\documentclass[acmsmall,screen,nonacm]{acmart}
\usepackage{ebproof}
\usepackage{mathtools}
\usepackage{ifthen}
\usepackage{natbib1}

\usepackage[capitalize,nameinlink]{cleveref}
\usepackage{tikz}
\usepackage{tikzpagenodes}
\usetikzlibrary{matrix,fit,quantikz,tikzmark}
\usepackage{pgfplots}
\usepgfplotslibrary{groupplots,colorbrewer}
\usepackage[inline]{enumitem}
\usepackage{moreenum}

\makeatletter
\patchcmd{\@addmarginpar}{\ifodd\c@page}{\ifodd\c@page\@tempcnta\m@ne}{}{}%
\makeatother
\reversemarginpar
\newif\ifcomments\commentsfalse
\ifcomments
    \advance\marginparwidth \marginparsep
    \advance\marginparwidth -4pt
    \advance\marginparwidth 210mm
    \advance\marginparwidth -6.75in
    \long\def\remark#1{
        \let\SAVEDspacefactor\relax
        \ifhmode\edef\SAVEDspacefactor{\the\spacefactor}\fi
        \raisebox{-3.5pt}[0pt][0pt]{\makebox[0pt][c]{$\scriptstyle\diamond$}}%
        \marginpar{\raggedright\hbadness=10000
        \parindent=8pt \parskip=2pt
        \def\baselinestretch{0.8}\footnotesize
        \itshape\noindent #1\par}%
        \ifx\SAVEDspacefactor\relax\else\spacefactor\SAVEDspacefactor\fi\relax}
\else
    \long\def\remark#1{\relax}%
\fi

\AtBeginDocument{%
  }

\setcopyright{acmlicensed}
\copyrightyear{2026}
\acmYear{2026}
\acmDOI{XXXXXXX.XXXXXXX}

\acmJournal{JACM}
\acmVolume{37}
\acmNumber{4}
\acmArticle{111}
\acmMonth{8}

\newcommand{\justH}{{\fboxsep=.3ex\fbox{$H$}}}

\DeclareMathAlphabet{\mathbbold}{U}{bbold}{m}{n}
\DeclareFontFamily{OT1}{pzc}{}
\DeclareFontShape{OT1}{pzc}{m}{it}{<-> s * [1.10] pzcmi7t}{}
\DeclareMathAlphabet{\mathpzc}{OT1}{pzc}{m}{it}

\newcommand{\data}[3]{\mathit{#1}_{\mathit{#2}}^{\mathit{#3}}}
\newcommand{\wmid}{\;\mid\;}
\newcommand{\kw}[1]{\textup{\textsf{#1}}}

\newcommand{\case}{}
\newcommand{\typeof}[2]{\Delta \vdash #1 : #2}

\newcommand{\denot}[1]{\llbracket{#1}\rrbracket}

\newcommand{\denotpv}[1]{\llbracket{#1}\rrbracket_{\mathrm{v}}}
\newcommand{\denotext}[1]{\llbracket{#1}\rrbracket_{\mathrm{e}}}
\newcommand{\denotpat}[1]{\llbracket{#1}\rrbracket_{\mathrm{p}}}

\newcommand{\Val}[1]{\mathcal{V}_{\mathit{#1}}}
\newcommand{\val}[1]{\mathpzc{v}_{\mathit{#1}}}

\newcommand{\mapping}[1]{\delta_{\mathit{#1}}}
\newcommand{\domctx}[1]{\textnormal{dom}(#1)}
\newcommand{\dom}[1]{\textnormal{funs}(#1)}
\newcommand{\tups}[1]{\textnormal{tups}(#1)}

\newcommand{\env}[1]{\mathit{Env}_{#1}}

\newcommand{\TyIsoTerm}[3]{#1 \vdash_{\mathrm{t}} #2 : #3}
\newcommand{\TyIsoIso}[3]{
  \ifthenelse{\equal{#1}{\varnothing}}{%
    \vdash_{\mathrm{i}} #2 : #3%
  }{%
    #1 \vdash_{\mathrm{i}} #2 : #3%
  }%
}
\newcommand{\TyIsoPattern}[3]{#1 \vdash_{\mathrm{p}} #2 : #3}
\newcommand{\TyIsoVal}[3]{#1 \vdash_{\mathrm{v}} #2 : #3}
\newcommand{\TyIsoExt}[3]{#1 \vdash_{\mathrm{e}} #2 : #3}

\newcommand{\TnIso}[1]{}

\newcommand{\odv}[2]{\text{OD}_{#1}(\{#2\})} 
\newcommand{\odvset}[2]{\text{OD}_{#1}(#2)} 
\newcommand{\odext}[2]{\text{OD}_{#1}^{\mathit{ext}}(\{#2\})}
\newcommand{\odextset}[2]{\text{OD}_{#1}^{\mathit{ext}}(#2)}

\newcommand{\fv}[1]{\mathit{fv}{(#1)}}
\newcommand{\botvalset}[1]{\mathit{BotVal*}{(#1)}}
\newcommand{\botval}[1]{\mathit{BotVal}{(#1)}}

\newcommand{\AddEnvP}[3]{\data{add}{p}{\delta}(#1, #2, #3)}
\newcommand{\AddEnvV}[3]{\data{add}{v}{\delta}(#1, #2, #3)}

\newcommand{\einsum}{\mathop{\mathsf{einsum}}}
\newcommand{\expand}{\mathop{\mathsf{expand}}}
\newcommand{\scatter}{\mathop{\mathsf{scatter}}}

\newcommand{\ValMatch}[2]{#1\ensuremath{\sim_{v}}#2}
\newcommand{\rw}[1]{{\texttt{\small\textbf{#1}}}}
\newcommand{\TT}{\texttt{tt}}
\newcommand{\FF}{\texttt{ff}}

\newcommand{\comp}[1]{\textit{comp}(#1)}
\newcommand{\comppat}[1]{\textit{comp$_{\mathrm{p}}$}(#1)}
\newcommand{\compv}[1]{\textit{comp$_{\mathrm{v}}$}(#1)}
\newcommand{\compext}[1]{\textit{comp$_{\mathrm{e}}$}(#1)}

\newcommand{\compenv}[1]{\textit{comp$_{\delta}$}(#1)}
\newcommand{\compsv}[2]{\textit{comp$_{\val{}}$}(#1, #2)}

\newcommand{\compet}[1]{\textit{comp$_{\mathrm{e}}^{\mathrm{t}}$}(#1)}
\newcommand{\compvt}[1]{\textit{comp$_{\mathrm{v}}^{\mathrm{t}}$}(#1)}
\newcommand{\comppt}[1]{\textit{comp$_{\mathrm{p}}^{\mathrm{t}}$}(#1)}
\newcommand{\fcompsv}{\textit{comp$_{\val{}}$}}

\newcommand{\disjoint}{\mathrel{\bot}}
\newcommand{\size}[2]{|#1|_{#2}}
\newcommand{\leftbox}[2]{%
  \tikzmark{#1}%
  \tikz[overlay,remember picture]{
    \node[
    inner sep=2pt,
    anchor=base west,
    rounded corners=2pt,
    font=\small,
    line width=1pt,
    gray!80!black,
    draw=gray!30,
    fill=gray!5,
    ] at (current page text area.west |- {pic cs:#1}) {#2};%
  }
}

\newcommand{\eqline}{\noalign{\smallskip} \noalign{\smallskip}}

\begin{document}

\title{Simulating Quantum Computation Using Probabilistic Programming}

\author{Yafei Yang}
\email{yafyang@iu.edu}
\orcid{0009-0006-1402-8014}
\author{Chung-chieh Shan}
\authornotemark[1]
\email{ccshan@iu.edu}
\affiliation{%
  \institution{Indiana University}
  \city{Bloomington}
  \state{Indiana}
  \country{USA}
}

\begin{abstract}
Developing quantum programs requires not only running them on quantum hardware but also simulating them on classical hardware.
Because the meaning of a quantum program can be defined in terms of state vectors, it~can be viewed as a form of weighted nondeterminism and simulated using any implementation of probabilistic programming that supports complex or at least negative weights.
In~this study, we~compile a high-level quantum programming language to one such implementation of probabilistic programming, a~solver for \emph{factor graph grammars}.
We~also extend the solver with an optimization that commutes the tensor operations $\einsum$ and $\expand$.
We~prove that the compiler and the optimization preserve the denotational semantics of the source language.
We perform benchmarks that show our straightforward implementation, by~repurposing existing infrastructure for probabilistic programming, attains simulation performance in the ballpark of several other quantum simulators.
\end{abstract}




\maketitle

\section{Introduction}

Quantum computations are still tricky to design today.
In order to understand and validate quantum programs, it is not enough to use quantum hardware to run them; we also need to use classical hardware to simulate them, even if we can only simulate toy problems.
Due to the inherent probabilistic behavior of quantum computation, simulation often involves predicting---either exactly or approximately---the probability distribution of outcomes of a computation, possibly as a function of some variable input or initialization.
The formal representation of the computation being simulated can be regarded as a quantum programming language.
For instance, a~quantum circuit is a program comprised of quantum gates operating on wires.

Typically, the semantics of quantum programs is defined by linear-algebra operations on vector spaces of quantum states.
For instance, a~quantum gate may denote a unitary linear transformation, and these denotations of individual quantum gates may be composed into the overall denotation of a program by additional operations such as tensor product.
Simulating a program amounts to computing its overall denotation.
However, denotations are vectors of high dimensionality (exponential in the number of qubits), so it can take too much time and memory for a simulator to naively represent them as arrays and iterate over them to compose them.
Instead, quantum simulators incorporate various optimizations, which require either special hardware or algorithms based on the qubit model \cite{ciceroSimulationQuantumComputers2025}.

An alternative view of quantum computation focuses on interpreting the exponentially many dimensions of quantum state vectors as a form of nondeterminism.
After all, it~has long been known that quantum computation involves ``negative probability'' \cite{feynmanSimulatingPhysicsComputers1982,muFunctionalQuantumProgramming2001}.
More recently, researchers have explored the use of probabilistic programming to simulate quantum correlations \cite{obeidEvaluatingProbabilisticProgramming2019} and understand quantum programming \cite{inoueQuantumProgrammingQuantum2024}.
Nevertheless, there seems to be no conclusion on how much probabilistic programming can simulate quantum computation.
Contemporaneously with quantum programming, many probabilistic programming languages have been designed in the past few decades \cite[among others]{koller-effective,van-de-meent-probabilistic,chiangExactRecursiveProbabilistic2023,moyRouletteLanguageExpressive2025}, along with optimized algorithms for probabilistic inference.
These developments lead to a natural question: can probabilistic programming help us simulate quantum computation?

In this paper, we show that yes, it~can.
This claim is supported by compiling a representative quantum language to a representative probabilistic language with little work and yielding competitive performance with other quantum simulators, instead of building a quantum simulator from scratch.
This approach has several motivations:
\begin{itemize}
\item Sharing techniques such as optimizations among implementations can help us understand and improve them.
\item An easy way to implement a classical simulator for a new quantum language may be to compile the language
to an existing probabilistic language.
\item The relationship between quantum and probabilistic programming deserves investigation when it comes to questions such as expressivity.
\end{itemize}

Our contributions are as follows.
\begin{enumerate}
\item We give the pure quantum programming language ISO \cite{sabrySymmetricPatternmatchingQuantum2018} an elementary denotational semantics that differs from the standard qubit and state vector model.
\item We formalize factor graph grammars (FGGs) \cite{chiangFactorGraphGrammars2020}, a graphical representation of probabilistic computation, as a language of tensor equations that includes $\einsum$, $\expand$, and $\scatter$ expressions. We give FGGs a denotational semantics that allows negative probability.
\item We compile ISO to FGGs.
\item We extend the existing FGG solver with a new optimization based on equation rewriting.
\item We prove the semantic correctness of our compiler and our new optimization.
\item We benchmark our and other simulators on a variety of quantum algorithms. These benchmarks show that our performance is in the ballpark of other simulators, thanks both to our new optimization and existing optimizations in the FGG solver. We hence demonstrate that quantum simulation can benefit from repurposing infrastructure for probabilistic programming.
\end{enumerate}

In~\cref{eg}, we give a motivating example that shows the similarity between quantum and probabilistic programming.
After defining the source language in \cref{src} and the target language in \cref{tgt}, we~show in \cref{comp} how to compile the source language to the target language while preserving denotational semantics.
In~\cref{opmt}, we show optimizations that build on the efficiency of the current FGG solver.
In~\cref{bench}, we describe benchmarks that measure the efficiency of our and other simulators.
We discuss related work in \cref{related} and future work in \cref{future}.

\section{Motivation}\label{eg}

We first motivate the design of a quantum programming language, then highlight the promise of compiling it to a probabilistic programming language.

\subsection{From Quantum Circuit to Higher-Level Quantum Programming}\label{sec:circ-qpl}

A standard way to describe a computation is to draw a circuit diagram.
For instance, the following circuit shows a ``fair quantum coin toss'':
\begin{equation}
\label{e:quantum-coin}
  \begin{quantikz}
    \lstick{\ket{0}} & \gate{H} & \qw
  \end{quantikz}
\end{equation}
This circuit initializes a qubit to the state~$\ket{0}$, then applies to~it a~Hadamard gate~$H$, which is one of the most fundamental quantum gates.
The result is a superposition of two states $\ket{0}$ and~$\ket{1}$.
It~is only when we measure this quantum coin that we observe one of the two possible outcomes (head and tail, so~to speak), each with probability~50\%.

The quantum circuit above precisely describes the operations we want to perform, but as a description, it~has three problems.
First, it~does not capture the intuition of a ``quantum coin'': there is no nondeterminism construct in~\labelcref{e:quantum-coin} where $\ket{1}$ appears alongside~$\ket{0}$.
Second, circuits relegate us to drawing wires that represent qubits and gates that operate on qubits. Qubits are convenient for a quantum computer, but tedious for humans when the circuit gets bigger or repetitive.
The third problem is there is no unified way to build a large circuit by composing parts already defined.
Except for standard gates like Hadamard \justH, there is no standard notation for one circuit to refer to another by name.

Fortunately, this is not the first time we encounter such problems.
Experiences from classical programming tell us that a well-designed programming language can often alleviate the burden for programmers.
For instance, pattern matching makes it possible to write clear conditions without the need of deeply nested if-statements.
Therefore, researchers introduce many classical ideas from programming languages to quantum programming, thus providing higher abstraction and freeing programmers from qubits and wires.

To see how the three problems above can be addressed, consider how our source language ISO expresses the same ``fair quantum coin toss'':
\begin{equation}\label{eg:had}
  \begin{aligned}
    &\rw{Hadamard} ::
      \mathbb{B} \leftrightarrow \mathbb{B}\\
    &\left\{
      \begin{aligned}
        &\FF \leftrightarrow
          \tfrac{1}{\sqrt{2}}\FF + \tfrac{1}{\sqrt{2}}\TT;\\
        &\TT \leftrightarrow
          \tfrac{1}{\sqrt{2}}\FF - \tfrac{1}{\sqrt{2}}\TT
      \end{aligned}
      \right\}\\[2ex]
    &\rw{Hadamard}~\FF&&
  \end{aligned}
\end{equation}
We define a function with the type $\mathbb{B} \leftrightarrow \mathbb{B}$.
The type $\mathbb{B}$ represents one qubit, whose two basic values \FF{} and \TT{} correspond to $\ket{0}$ and $\ket{1}$ in popular notation.
The bidirectional arrow~$\leftrightarrow$ indicates that this function is \emph{reversible}: it~can be applied either left-to-right or right-to-left.
The last line of~\labelcref{eg:had} applies \rw{Hadamard} left-to-right:
Because the input $\FF$ matches the left hand side of the first clause, the output is the linear combination $\tfrac{1}{\sqrt{2}}\FF + \tfrac{1}{\sqrt{2}}\TT$ on the right hand side.
For such a~superposition state, the squares of the coefficients in front of basic values must add up to 1.
If~we were to apply \rw{Hadamard} to $\FF$ right-to-left instead, then the \emph{right} sides of \emph{both} clauses would match \emph{fractionally}, to produce the corresponding linear combination $\tfrac{1}{\sqrt{2}}\FF + \tfrac{1}{\sqrt{2}}\TT$ of the left hand sides.

Even though this representation of the Hadamard gate looks much longer than the circuit~\justH{} or the matrix $\tfrac{1}{\sqrt{2}}\bigl[\begin{smallmatrix}1&1\\1&-1\end{smallmatrix}\bigr]$ at first glance, it captures the intuition of a quantum coin by linear combinations of basic values.
Since we use types like $\mathbb{B}$, we can avoid mentioning qubits explicitly.
Thus, we can replace the type $\mathbb{B}$ with a different one and represent a reversible gate on multiple qubits without the need of drawing all of them explicitly, which reduces the work tremendously when the circuit size becomes larger.
For instance, if we have a type $a$, we can use $a \oplus a$ to represent the union type whose left and right values are both of type $a$.
Suppose $\kw{inj}_{l}~\_$ and $\kw{inj}_{r}~\_$ construct the left and right value of a union type respectively. We~can define the following function:
\begin{equation}\label{eg:had*}
  \begin{aligned}
    &\rw{Hadamard*} ::
      a \oplus a \leftrightarrow a \oplus a\\
    &\left\{
      \begin{aligned}
        &\kw{inj}_{l}~x \leftrightarrow
          \tfrac{1}{\sqrt{2}}(\kw{inj}_{l}~x) + \tfrac{1}{\sqrt{2}}(\kw{inj}_{r}~x);\\
        &\kw{inj}_{r}~x \leftrightarrow
          \tfrac{1}{\sqrt{2}}(\kw{inj}_{l}~x) - \tfrac{1}{\sqrt{2}}(\kw{inj}_{r}~x)
      \end{aligned}
      \right\}
  \end{aligned}
\end{equation}
Once we have these function definitions, we can invoke them later in new definitions.
Thus, we get a larger quantum program by composing smaller ones.

\subsection{Simulating Quantum Programming with Probabilistic Programming}

Well-designed programming languages like above solve all three problems with the circuit model.
But writing nice programs is not enough.
We also want to run the programs.
Since quantum computers are not widely available yet, we want to simulate the programs on a classical computer.
Even without the consideration of optimization, implementing such a simulator already takes a lot of work \cite{ciceroSimulationQuantumComputers2025}.
Worse, a naive implementation can only simulate very few qubits and gates, thus making it infeasible.

Consider the programs \rw{Hadamard} and \rw{Hadamard*} in \cref{sec:circ-qpl}.
Whereas \rw{Hadamard} denotes a $2 \times 2$ matrix $\tfrac{1}{\sqrt{2}}\left[\begin{smallmatrix}1&1\\1&-1\end{smallmatrix}\right]$, the more general program \rw{Hadamard*} denotes a $2^{n+1} \times 2^{n+1}$ matrix
\begin{equation}
\frac{1}{\sqrt{2}}
\begin{tikzpicture}[baseline=(m.center), every label/.style={font=\footnotesize}, label distance=1ex]
    \matrix (m) [matrix of math nodes, ampersand replacement=\&,
                 left delimiter={[}, right delimiter={]}, inner xsep=0,
                 every node/.append style={inner sep=0}] {
        \hphantom{-}1 \& \& \&[1ex] \hphantom{-}1 \& \& \vphantom{-1} \\[-1ex]
        \& \ddots \& \& \& \ddots \& \\
        \& \& 1\hphantom{-} \& \& \& 1\hphantom{-} \\[1ex]
        \hphantom{-}1 \& \& \& -1 \& \& \vphantom{-1} \\[-1ex]
        \& \ddots \& \& \& \ddots \& \\
        \hphantom{-1} \& \& 1\hphantom{-} \& \hphantom{-1} \& \& -1 \\
    };
    \node [fit=(m-6-1) (m-6-3), below delimiter=\}, inner sep=0, label=below:$2^n$] {};
    \node [fit=(m-6-4) (m-6-6), below delimiter=\}, inner sep=0, label=below:$2^n$] {};
    \begin{scope}[transform canvas={xshift=.5em}]
        \node [fit=(m-1-6) (m-3-6), right delimiter=\}, inner sep=0, label=right:$2^n$] {};
        \node [fit=(m-4-6) (m-6-6), right delimiter=\}, inner sep=0, label=right:$2^n$] {};
    \end{scope}
\end{tikzpicture}
\end{equation}
where $n$ is the number of qubits required by the type~$a$.
A~naive implementation would use $4^{n+1}$ locations to store all numbers in the matrix.
Because of the exponential increase of memory requirement, a naive simulator cannot scale very well.
But if we observe the structure of \rw{Hadamard*}, we can see that it can be split into four $2^{n} \times 2^{n}$ blocks, each a scalar multiple of an identity matrix.
Thus, we only need to store the size of each block and their coefficients, saving a lot of space.

This observation is not unique to this quantum computation, or to quantum programming.
Probabilistic programming has been using inference algorithms that can be implemented as efficient tensor operations for a long time \cite{kschischangFactorGraphsSumproduct2001,poonSumproductNetworksNew2012,chiangImplementation2025}.
The implementations of these algorithms, such as sum-product networks, don't assume that the probabilities have to be positive or even real numbers.
Also, the linear combinations in~\labelcref{eg:had,eg:had*} above resemble the ambiguous operator \kw{amb} in probabilistic programming \cite{DBLP:conf/aieeire/000161,chiangExactRecursiveProbabilistic2023}.
The similarity in both implementation and syntax makes it promising to simulate quantum programming with probabilistic programming.
If we have a new quantum language, it should be easy to simulate it by compiling it to a probabilistic language that already has efficient inference algorithms.
Then we get a quantum simulator almost for free.

In this paper, we use \emph{factor graph grammars} (FGGs) \cite{chiangFactorGraphGrammars2020} as an instance of a probabilistic language that can serve as a compilation target.
In the next few sections, we introduce in detail how to compile the pure quantum programming language ISO to FGGs.

\section{Source Language: ISO}\label{src}

Recall that in order to justify our main claim---that probabilistic programming can help us simulate quantum computation---we compile a quantum language to a probabilistic language.
To~formulate this compilation, we require that the source language satisfy two criteria.
\begin{enumerate}[leftmargin=*,labelindent=0pt]
\item First, we require that the source language expresses exactly terminating pure quantum programs.
We eliminate non-terminating programs because most quantum algorithms terminate, and the meaning of non-terminating quantum programs is still ongoing research \cite{assoliniDenotationalSemanticsQuantum2025}.
We eliminate non-pure programs, i.e., programs that have measurements and so are irreversible, because reversibility is a key postulate for quantum computation \cite{nielsenQuantumComputationQuantum2010}.
\item Second, we require that the source language have a clear semantics.
This way, to prove that the simulation does capture the meaning of a quantum program, we just need to prove that the compilation from quantum programs to probabilistic programs preserves the semantics.
\end{enumerate}

Any quantum programming language satisfying the two criteria above can serve our purpose.
Here, we choose ISO \cite{sabrySymmetricPatternmatchingQuantum2018} as a representative quantum programming language.
ISO satisfies the two criteria above.
\begin{enumerate}[leftmargin=*,labelindent=0pt]
\item First, it is straightforward to pick a subset of ISO that expresses exactly terminating pure quantum programs.
For the terminating part, ISO can be simplified or extended to serve different purposes easily.
In fact, ISO has been tailored to demonstrate both the Curry-Howard correspondence between quantum programs and proofs as well as the Turing completeness of quantum computing \cite{chardonnetCurryHowardCorrespondenceQuantum2023}.
Thus, we can choose a subset of ISO that allows only terminating programs.
For the pure part, the support of reversibility is at the core of ISO\@.
One of its main features, \emph{symmetric pattern matching}, behaves like isomorphisms.
This feature naturally models unitary gates in quantum computing.
\item Second, ISO's type system benefits our presentation of denotational semantics and correctness proof.
Because ISO abstracts away qubits by algebraic data types, we can give it a clear denotational semantics, which in turn makes the correctness proof straightforward.
\end{enumerate}
We give the syntax, the typing system, and the denotational semantics of ISO below.

\subsection{Syntax}

\begin{figure}
\begin{alignat*}{4}
    &\text{Programs} && (\data{Prog}{iso}{})
    & \mathit{prog} & ::=\; && \omega \wmid t \\
    &\text{Isos} && (\data{Iso}{iso}{})
    & \omega & ::= && \{~
                      \case v_{1} \leftrightarrow e_{1}
                      \cdots
                      \case v_{n} \leftrightarrow e_{n}~
                      \}\\
    &\text{Terms} && (\data{Term}{iso}{})
    & t   & ::= && \kw{()} \wmid x \wmid \kw{inj}_{l}~t \wmid \kw{inj}_{r}~t
                   \wmid \langle{t_{1}, t_{2}}\rangle \wmid \omega~t \wmid \\
    & && & & && \kw{let}~p~\kw{=}~t_{1}~\kw{in}~t_{2}
                \wmid t_{1} + t_{2}
                \wmid \alpha{t}\\
    &\text{Extended values} && (\data{ExtValue}{iso}{})
    & e   & ::= && v \wmid e_{1} + e_{2} \wmid \alpha{e} \wmid
                   \kw{let}~p_{1}~\kw{=}~\omega~{p_{2}}~\kw{in}~e\\
    &\text{Pure values} && (\data{Value}{iso}{})
    & v   & ::= && \kw{()} \wmid x \wmid \kw{inj}_{l}~v \wmid \kw{inj}_{r}~v
                   \wmid \langle{v_1, v_2}\rangle \\
    &\text{Patterns} && (\data{Pattern}{iso}{})
    & p   & ::= && \kw{()} \wmid x \wmid \langle{p_{1},p_{2}}\rangle \\
    &\text{Value and term variables} &&& x \\
    &\text{Weights} &&
    & \alpha & \in \mathbb{C} \hidewidth && \\
    &\text{Value types} && (\data{Type}{iso}{val})
    & a,b & ::= && \mathbbold{1} \wmid a \otimes b \wmid a \oplus b\\
    &\text{Iso types} && (\data{Type}{iso}{iso})
    & T   & ::= && a \leftrightarrow b\\
\end{alignat*}
\caption{The syntax of ISO.}
\label{fig:iso-syntax}
\end{figure}

The syntax of ISO is shown in \cref{fig:iso-syntax}.
In ISO\@, a program is either an iso~$\omega$ or a term~$t$.

An~iso consists of a collection of clauses $v \leftrightarrow e$, each with a pair of pure value~$v$ on the left and an extended value~$e$ on the right.
Extended values extend pure values with combination ($e_{1} + e_{2}$), scale ($\alpha{e}$), and local binding ($\kw{let}$).
Thus, an iso has the syntax of pattern matching, and it represents a unitary operation.

A term is similar to an extended value.
The difference is that in a \kw{let} term, the right hand side can be any other term, while in an extended value, the right hand side has to be $\omega~p$.
This requirement makes it easy to turn each iso term into another iso term that represents the inverse of the unitary operation.
Pure values, extended values, and patterns are all subsets of terms.

The weights~$\alpha$ in extended values and terms range over complex numbers.

With this syntax, \rw{Hadamard} in \cref{eg:had} can be written as follows:
\begin{equation}\label{eg:iso-had}
  \begin{aligned}
    &\rw{Hadamard} ::
      \mathbbold{1} \oplus \mathbbold{1} \leftrightarrow \mathbbold{1} \oplus \mathbbold{1}\\
    &\left\{
      \begin{aligned}
        &\mid \kw{inj}_{l} () \leftrightarrow
          \tfrac{1}{\sqrt{2}}\kw{inj}_{l} () + \tfrac{1}{\sqrt{2}}\kw{inj}_{r} () \\
        &\mid \kw{inj}_{r} () \leftrightarrow
          \tfrac{1}{\sqrt{2}}\kw{inj}_{l} () - \tfrac{1}{\sqrt{2}}\kw{inj}_{r} ()
      \end{aligned}
      \right\}\\[2ex]
    &\rw{Hadamard}~\kw{inj}_{l} ()&&
  \end{aligned}
\end{equation}

\subsection{Type System}

\begin{figure}
\begin{alignat*}{4}
    &\text{Typing contexts} &&& \Delta & ::=\; && \varnothing \wmid \Delta,x:a \\
    &\text{Iso types} && (\data{Type}{iso}{iso})
    & T   & ::= && a \leftrightarrow b\\
    &\text{Value types} && (\data{Type}{iso}{val}) \quad
    & a,b & ::= && \mathbbold{1} \wmid a \oplus b \wmid a \otimes b
\end{alignat*}
\flushleft{\fbox{$\TyIsoTerm{\Delta}{t}{a}$}}
\begin{gather*}
{
\begin{prooftree}[center=false]
  \hypo{}
  \infer1[\TnIso{UNIT$_{t}$}]{%
    \TyIsoTerm{\varnothing}%
    {\kw{()}}%
    {\mathbbold{1}}}
\end{prooftree}
}
\quad\quad
{
\begin{prooftree}[center=false]
  \hypo{}
  \infer1[\TnIso{VAR$_{t}$}]{\TyIsoTerm{x : a}{x}{a}}
\end{prooftree}
}
\quad\quad
{
\begin{prooftree}[center=false]
  \hypo{\TyIsoTerm{\Delta}{t}{a}}
  \infer1[\TnIso{INJL$_{t}$}]{\TyIsoTerm{\Delta}{\kw{inj}_{l}~t}{a \oplus b}}
\end{prooftree}
}
\quad\quad
{
\begin{prooftree}[center=false]
  \hypo{\TyIsoTerm{\Delta}{t}{b}}
  \infer1[\TnIso{INJR$_{t}$}]{\TyIsoTerm{\Delta}{\kw{inj}_{r}~t}{a \oplus b}}
\end{prooftree}
}
\\
{
\begin{prooftree}[center=false]
  \hypo{\TyIsoTerm{\Delta_{1}}{t_{1}}{a}}
  \hypo{\TyIsoTerm{\Delta_{2}}{t_{2}}{b}}
  \infer2[\TnIso{PAIR$_{t}$}]{\TyIsoTerm{\Delta_{1};\Delta_{2}}{\langle{t_{1}, t_{2}}\rangle}{a \otimes b}}
\end{prooftree}
}
\quad\quad
{
\begin{prooftree}[center=false]
  \hypo{\TyIsoIso{\varnothing}{\omega}{a \leftrightarrow b}}
  \hypo{\TyIsoTerm{\Delta}{t}{a}}
  \infer2[\TnIso{APP$_{t}$}]{\TyIsoTerm{\Delta}{\omega~t}{b}}
\end{prooftree}
}
\\
{
\begin{prooftree}[center=false]
  \hypo{\TyIsoTerm{\Delta_{1}}{t_{1}}{a}}
  \hypo{\TyIsoPattern{\Delta_{2}'}{p}{a}}
  \hypo{\TyIsoTerm{\Delta_{2};\Delta_{2}'}{t_{2}}{b}}
  \infer3[\TnIso{LET$_{t}$}]{\TyIsoTerm{\Delta_{1};\Delta_{2}}{\kw{let}~p~\kw{=}~t_{1}~\kw{in}~t_{2}}{b}}
\end{prooftree}
}
\quad\quad
{
\begin{prooftree}[center=false]
  \hypo{\TyIsoTerm{\Delta}{t_{1}}{a}}
  \hypo{\TyIsoTerm{\Delta}{t_{2}}{a}}
  \infer2[\TnIso{SUM$_{t}$}]{\TyIsoTerm{\Delta}{t_{1} + t_{2}}{a}}
\end{prooftree}
}
\quad\quad
{
\begin{prooftree}[center=false]
  \hypo{\TyIsoTerm{\Delta}{t}{a}}
  \infer1[\TnIso{SCALE$_{t}$}]{\TyIsoTerm{\Delta}{\alpha{t}}{a}}
\end{prooftree}
}
\end{gather*}
\flushleft{\fbox{$\TyIsoIso{\varnothing}{\omega}{T}$}}
\begin{gather*}
{
\begin{prooftree}[center=false]
  \hypo{
    {\begin{matrix*}[l]
      {\TyIsoVal{\Delta}{v_{1}}{a}}
      &{\dots}
      &{\TyIsoVal{\Delta}{v_{n}}{a}}
      &\odv{a}{v_{1}, \dots, v_{n}}\\
      &\\
      {\TyIsoExt{\Delta}{e_{1}}{b}}
      &{\dots}
      &{\TyIsoExt{\Delta}{e_{n}}{b}}
      &\odext{b}{e_{1}, \dots, e_{n}}\\
    \end{matrix*}}
  }
  \hypo{
    {\begin{bmatrix}
      \alpha_{11} & \cdots & \alpha_{1n}\\
      \vdots & & \vdots\\
      \alpha_{n1} & \cdots & \alpha_{nn}\\
    \end{bmatrix}~\text{is unitary}}
  }
  \infer2[\TnIso{ISO}]{%
    \TyIsoIso{\varnothing}
    {\begin{Bmatrix*}[l]
      \case v_{1} \leftrightarrow \alpha_{11}e_{1} + \cdots + \alpha_{1n}e_{n}\\
      \dots\\
      \case v_{n} \leftrightarrow \alpha_{n1}e_{1} + \cdots + \alpha_{nn}e_{n}~\\
    \end{Bmatrix*}%
  }%
  {a \leftrightarrow b}}
\end{prooftree}
}
\end{gather*}
\caption{The type system of ISO (abridged)}
\label{fig:iso-type}
\end{figure}

ISO uses a standard linear type system, shown in \cref{fig:iso-type}.
We~assume that a typing context~$\Delta$ has no duplicate variable names.
We~write $\Delta_{1};\Delta_{2}$ to concatenate two typing contexts.

For every nonterminal in the syntax, we have a typing judgment form.
For terms, we have typing judgments of the form $\TyIsoTerm{\Delta}{t}{a}$, where $\Delta$ is a typing context, $t$ is a term, and $a$ is a value type.
Since pure values, extended values, and patterns are all subsets of terms, the typing judgments of pure values $\TyIsoVal{\Delta}{v}{a}$, extended values $\TyIsoExt{\Delta}{e}{a}$, and patterns $\TyIsoPattern{\Delta}{p}{a}$ are defined in a similar way and omitted here.

Based on the typing rules of pure values, extended values, and patterns, we have the typing rule of isos at the bottom of \cref{fig:iso-type}.
The constraint $\odv{a}{v_{1},\dotsc,v_{n}}$ takes a value type $a$ and a set of pure values $\{v_{1},\cdots,v_{n}\}$ of type $a$.
It guarantees that $v_{1}, \dotsc, v_{n}$ cover all possible values of type $a$ (\emph{exhaustive}) and no two $v_i, v_j$ match the same value (\emph{non-overlapping}).
The constraint $\text{OD}^{\mathit{ext}}$ does the same work on extended values.
The two constraints and the unitary side-condition together ensure that an iso is reversible and thus implementable as a quantum gate.
Once a program satisfies these constraints, we can ignore them in the compilation and simulation, whose correctness does not depend on the constraints. Therefore, we omit the definitions of the constraints.
We refer readers to \citet{sabrySymmetricPatternmatchingQuantum2018} for the details of $\text{OD}$ and $\text{OD}^{\mathit{ext}}$.

\subsection{Denotational Semantics}\label{sec:iso-denot}

There have been a lot of attempts to give a quantum programming language a denotational semantics \cite{selingerQuantumProgrammingLanguage2004,choSemanticsQuantumProgramming2014,valironSemanticsQuantumProgramming2022,assoliniDenotationalSemanticsQuantum2025}.
ISO and its variants also have different denotational semantics based on Hilbert spaces \cite{sabrySymmetricPatternmatchingQuantum2018} or category theory \cite{chardonnetSemanticsTuringcompleteReversible2023,daveCombiningQuantumClassical2025}.
Instead of emphasizing the reversible behavior of quantum programs and using linear operators in Hilbert spaces, here we focus on the left-to-right direction of an iso and think of it as the quantum version of a conditional probability distribution. 
To~define this denotational semantics of ISO\@, we first define semantic values of ISO\@:
\begin{equation*}
  \begin{aligned}
    &\text{Semantic Values} && (\Val{I})
    & \Val{I}   & ::= && \kw{()} \wmid
                         \langle{\Val{I_{1}}, \Val{I_{2}}}\rangle \wmid
                         \kw{inj}_{l}~\Val{I} \wmid
                         \kw{inj}_{r}~\Val{I} 
  \end{aligned}
\end{equation*}
For any type $a$, its semantic values can then be thought to form a basis of a Hilbert space.

Types in ISO also have their semantics.
The semantics of a type~$a$ in ISO is defined as a set~$\denot{a}$ of semantic values:
\begin{alignat}{2}
    &\denot{\mathbbold{1}} &
    &=\{\kw{()}\}\\
    &\denot{a \oplus b} &
    &=\{\kw{inj}_{l}~v \mid v \in \denot{a}\} \cup \{\kw{inj}_{r}~v \mid v \in \denot{b}\}\\
    &\denot{a \otimes b} &
    &=\{\langle{v_{1}, v_{2}}\rangle \mid v_{1} \in \denot{a}, v_{2} \in \denot{b}\}
\end{alignat}

Based on semantic values, we can define ISO environments:
\begin{definition}[ISO Environment]
  Given a typing context~$\Delta$, a corresponding environment $\mapping{}$ is a mapping that maps each variable $x \in \domctx{\Delta}$ to a semantic value of type $\Delta(x)$.
  We use $\denot{\Delta}$ to represent the set of environments that correspond to a typing context $\Delta$.
\end{definition}
We write $\env{I}$ to represent the collection of ISO environments.
We use $\varnothing$ to represent an environment whose domain is empty, and $\{(x,\val{})\}$ to represent a singleton environment that maps one variable~$x$ to~$\val{}$.

A term denotes a function from an environment and a semantic value to a weight.
\Crefrange{denot:iso-t-unit}{denot:iso-t-scale} define this denotation by induction on the typing judgment.
The denotational semantics of patterns, pure values, and extended values are defined in a similar way and can be found in Appendix~\ref{section:denot-iso-more}.
{
\allowdisplaybreaks
\begin{alignat}{2}
  \label{denot:iso-t-unit}
  &\denot{\TyIsoTerm{\varnothing}{\kw{()}}{\mathbbold{1}}}(\varnothing, \kw{()})
  &&=
  1\\
  &\denot{\TyIsoTerm{x:a}{x}{a}}(\{(x,\val{}')\}, \val{})
  &&=
  \begin{cases}
    1 & \text{if}~\val{}' = \val{}\\
    0 & \text{otherwise}
  \end{cases}\\
  &\denot{\TyIsoTerm{\Delta}{\kw{inj}_{l}~t}{a \oplus b}}(\delta, \kw{inj}_{i}~\val{})
  &&=
  \begin{cases}
    \denot{\TyIsoTerm{\Delta}{t}{a}}(\delta, \val{}) & \text{if}~i=l \\
    0 & \text{if}~i=r \\
  \end{cases}\\
  &\denot{\TyIsoTerm{\Delta}{\kw{inj}_{r}~t}{a \oplus b}}(\delta, \kw{inj}_{i}~\val{})
  &&=~
  \begin{cases}
    \denot{\TyIsoTerm{\Delta}{t}{b}}(\delta, \val{}) & \text{if}~i=r \\
    0 & \text{if}~i=l \\
  \end{cases}\\
  &\denot{\TyIsoTerm{\Delta_{1};\Delta_{2}}%
    {\langle{t_{1}, t_{2}}\rangle}%
    {a \otimes b}}(\delta_{1} \cup \delta_{2}, \langle{\val{1}, \val{2}}\rangle)
  &&=
  \denot{\TyIsoTerm{\Delta_{1}}%
    {t_{1}}%
    {a}}(\delta_{1}, \val{1}) \cdot
    \denot{\TyIsoTerm{\Delta_{2}}%
    {t_{2}}%
    {b}}(\delta_{2}, \val{2})\\
  \label{denot:iso-t-app}
  &\denot{\TyIsoTerm{\Delta}{\omega~t}{b}}(\delta, \val{}')
  &&=
    \smash[b]{\sum_{\val{} \in \denot{a}}}%
    \denot{%
    \TyIsoIso{\varnothing}%
    {\omega}%
    {a \leftrightarrow b}}%
    (\val{}, \val{}') \cdot
    \denot{\TyIsoTerm{\Delta}{t}{a}}(\delta, \val{})\\
  \label{denot:iso-t-let}
  &\denot{\TyIsoTerm{\Delta_{1};\Delta_{2}}{\kw{let}~p~\kw{=}~t_{1}~\kw{in}~t_{2}}{b}}%
    (\delta_{1} \cup \delta_{2}, \val{}')\notag\\
  &\quad=\mathrlap{%
    \sum_{\delta_{2}' \in \denot{\Delta_{2}'}}
    \sum_{\val{} \in \denot{a}}
      \denot{\TyIsoTerm{\Delta_{1}}{t_{1}}{a}}%
        (\delta_{1}, \val{}) \cdot
      \denotpat{\TyIsoPattern{\Delta_{2}'}{p}{a}}%
        (\delta_{2}', \val{}) \cdot
        \denot{\TyIsoTerm{\Delta_{2};\Delta_{2}'}{t_{2}}{b}}%
        (\delta_{2} \cup \delta_{2}', \val{}')
    }\\
  &\denot{\TyIsoTerm{\Delta}{t_{1} + t_{2}}{a}}(\delta, \val{})
  &&=
  \denot{\TyIsoTerm{\Delta}{t_{1}}{a}}(\delta, \val{}) +
    \denot{\TyIsoTerm{\Delta}{t_{2}}{a}}(\delta, \val{})\\
  \label{denot:iso-t-scale}
  &\denot{\TyIsoTerm{\Delta}{\alpha{t}}{a}}(\delta, \val{})
  &&=
  \alpha \cdot \denot{\TyIsoTerm{\Delta}{t}{a}}(\delta, \val{})
\end{alignat}
}

An iso denotes a function from a pair of semantic values to a weight:
\begin{gather}
  \label{denot:iso-iso}
  \denot{\TyIsoIso{\varnothing}
    {\{~
      \case v_{1} \leftrightarrow e_1
      \cdots
      \case v_{n} \leftrightarrow e_n~
    \}}{a \leftrightarrow b}}
    (\val{}, \val{}')
  \qquad\notag\\\qquad\qquad=
    \sum_{i = 1}^{n}%
    \sum\limits_{\delta \in \denot{\Delta}}
    \denot{\TyIsoVal{\Delta}%
    {v_i}%
    {a}}%
    (\delta, \val{})\cdot
    \denot{\TyIsoExt{\Delta}%
    {e_i}%
    {b}}%
    (\delta, \val{}')
\end{gather}

\section{Target Language: Factor Graph Grammars}\label{tgt}

With the source language at hand, we can now consider the criteria for the target probabilistic language.
We consider four criteria.
\begin{enumerate}[leftmargin=*,labelindent=0pt]
\item First, the target language should be general enough to allow negative or complex weights.
  Without such generality, representing quantum amplitudes faithfully becomes difficult.
\item Second, the target language should have a clear semantics.
  Without such a semantics, we cannot guarantee that the meanings of source programs and target programs match.
  In this case, even if we are able to run the simulation, we cannot guarantee correctness.
\item Third, the target language should have an efficient solver.
  After all, our aim is to show that infrastructure of probabilistic programming can be repurposed for quantum simulation.
  If the target language doesn't have a solver, there is little to reuse.
\item Fourth, the existing solver should be easy to extend and optimize.
  This makes it possible to obtain a quantum simulator from the existing solver with minimal effort.
\end{enumerate}

Any probabilistic programming language satisfying the four criteria above can serve our purpose.
Here, we choose \emph{factor graph grammars} (FGGs) \cite{chiangFactorGraphGrammars2020} as a representative probability model.
As the name implies, an FGG model is a grammar that generates a set of graphs, and the weight of each graph is a product of factors.
FGGs satisfy the four criteria above.
\begin{enumerate}[leftmargin=*,labelindent=0pt]
\item First, FGGs can be generalized to any semiring and thus support complex weights.
In FGGs as defined by \citet{chiangFactorGraphGrammars2020}, the weights are non-negative and the operations are monotone, so even recursive grammars admit least-fixed-point solutions.
But here, we do not consider recursive quantum programs or quantum loops \cite{selingerQuantumProgrammingLanguage2004}, so the weights can be negative or even complex.
\item Second, we can give FGGs a clear semantics by representing them as a finite set of tensor equations instead of a graph grammar.
Though the two representations are equivalent, the former representation leads to a semantics closer to the ISO semantics in \cref{sec:iso-denot}, which makes the semantic preservation proof straightforward.
In this representation, the left hand side of each equation is a tensor variable, and the right hand side of each equation is built out of tensor variables and operations.
For instance, the following equations compute the probability of getting the same result from two flips of the same unfair coin:
\begin{align}
\label{e:fgg-example}
    \textrm{coin} &= \tfrac{2}{3}\cdot[1\,0] + \tfrac{1}{3}\cdot[0\,1]
&
    \textrm{same} &= \textrm{coin} \cdot \left[\begin{smallmatrix}1&0\\0&1\end{smallmatrix}\right] \cdot \textrm{coin}^t
\end{align}
\item Third, FGGs already have an efficient solver \cite{chiangExactRecursiveProbabilistic2023}.
For instance, solving \cref{e:fgg-example} gives us
\begin{align}
    \textrm{coin} &= [\tfrac{2}{3}\,\tfrac{1}{3}]
&
    \textrm{same} &= \tfrac{5}{9}.
\end{align}
\item Fourth, this solver is open source and turns out to be easy to extend and modify.
Since its implementation is modular and doesn't rely on the assumption that weights must be non-negative, it is also straightforward to add support for complex weights.
\end{enumerate}
We give the syntax, the denotational semantics, and the sparse tensor representation of FGGs below.

\subsection{Syntax}\label{section:fgg-syntax}

\begin{figure}
\begin{equation*}
\begin{aligned}
    &\text{FGGs} && (\data{Prog}{fgg}{}) & \mathit{fgg} &::={} && (t=E)\dotso \\
    &\text{Tensor expressions} && (\data{Tensor}{}{}) & E,F &::={} &&t \wmid \alpha \wmid E+E \wmid 0^{(s,\dotsc)} \wmid \\
    &&&&&&&\expand(E[\vec{x}] \to \vec{z}) \wmid \\
    &&&&&&&\einsum(E[\vec{x}],\dotsc \to \vec{z}) \wmid \\
    &&&&&&&\scatter(E[\vec{x}] \to e\dotsm) \\
    &\text{Tensor variables} && (\data{TVar}{}{})  & t \\
    &\text{Index variable sequences}  && (\data{IVars}{}{}) & \vec{x},\vec{y},\vec{z} &::={} && x^{(s)}\dotsm \\
    &\text{Axis expressions} && (\data{Axis}{}{}) & e &::={} && x^{(s)} \wmid () \wmid s\oplus e\oplus s \wmid e\otimes e \\
    &\text{Index variables} && (\data{IVar}{}{}) & x^{(s)},y^{(s)},z^{(s)} \\
    &\text{Upper bounds} && & s & \in \mathbb{N}
\end{aligned}
\end{equation*}
\caption{The syntax of FGGs, expressed as a system of tensor equations.}
\label{fig:fgg-syntax}
\end{figure}

The syntax of FGGs is shown in \cref{fig:fgg-syntax}.

A tensor equation $t=E$ defines the tensor variable~$t$ by a tensor expression~$E$.
Each tensor variable~$t$ will turn out to denote a multidimensional array of numbers.
The first four cases of tensor expressions are simple: a~tensor can be a variable, a literal scalar~$\alpha$, the sum of two tensors, or the zero tensor of a specified shape $(s,\dotsc)$, where each $s$ is a natural number.

The last three cases of tensor expressions---$\expand$, $\einsum$, and $\scatter$---sport more elaborate syntax.
They transform the shape of tensors.
The $\expand$ operation adds dimensions to a tensor.
The $\einsum$ operation generalizes scalar, vector, matrix, and elementwise multiplication.
The $\scatter$ operation transforms a dense tensor to a sparse tensor.

To specify the shapes of input and output for each operation, we use \emph{index variable sequences}.
For instance, $\expand$ can transform the vector $[6\,7]$ to $[\begin{smallmatrix}6&6\\7&7\end{smallmatrix}]$
\begin{equation}
  \expand([6\,7][x^{(2)}] \to x^{(2)}y^{(2)}) = [\begin{smallmatrix}6&6\\7&7\end{smallmatrix}]
  \label{eg:expand-67}
\end{equation}
In this expression, an index variable sequence $x^{(2)}$ is used to specify the shape of the input tensor, and another index variable sequence $x^{(2)}y^{(2)}$ is used to specify the shape of the output tensor.
Each index variable sequence consists of a sequence of \emph{index variables}.
Since all three tensor operations are based on index variables, below we first introduce index variables, then the three operations.

\subsubsection{Index Variables}
An \emph{index variable}~$x^{(s)}$ specifies which tensor dimensions to operate on.
For instance, in $\expand([6\,7][x^{(2)}] \to x^{(2)}y^{(2)})$ from \cref{eg:expand-67}, the variable $x^{(2)}$ specifies the only dimension of the input tensor $[6\,7]$ to be of size $2$ and name it as $x$.
The variable $y^{(2)}$ specifies the dimension that we want to add so that $\expand$ knows where to do the expansion.
Since an index variable has an \emph{upper bound}~$^{(s)}$, it iterates over integers between $0$ and $s-1$.
We often elide the upper bound~$^{(s)}$ even though it is formally present.
We write $\vec{x}$ or $\vec{y}$ or $\vec{z}$ for an \emph{index variable sequence}.
We notate concatenation of sequences by juxtaposition:~$\vec{x}\vec{y}$.
We define the union sequence $\vec{x}\cup\vec{y}$ to be the concatenation $\vec{x}\vec{y}$ with duplicates removed (leaving the first occurrence of each index variable, say, so $xy \cup xzx = xyz$).
The kind of equational reasoning we perform below lets us reorder tensor dimensions and corresponding sequences without loss of generality \citep{chiang2023named}.

Index variables in different operations are interpreted in separate scopes.
For instance, if we nest $\expand$ from \cref{eg:expand-67} into another $\expand$ as below:
\begin{equation}
  \expand(\expand([6\,7][x^{(2)}] \to x^{(2)}y^{(2)})[z^{(2)}w^{(2)}] \to \underline{x}^{(2)}z^{(2)}w^{(2)})
\end{equation}
The underlined $\underline{x}$ is interpreted as a new dimension added to the input of the outer $\expand$, rather than referring to the one without underline from the inner $\expand$.
Depending on the operation, index variables in an operation can be used different numbers of times.
These specific requirements are introduced below.

Two index variable sequences $\vec{x}, \vec{y}$ are \emph{disjoint} if they share no index variable, which is written as $\vec{x} \disjoint \vec{y}$.
An index variable sequence $\vec{x}$ is a \emph{subset} of $\vec{y}$ if every index variable occurring in $\vec{x}$ also occurs in $\vec{y}$, which is written as $\vec{x} \sqsubseteq \vec{y}$.
We use these relations in the proof of \cref{thm:einsum-expand}.

\subsubsection{$\expand$}
The~$\expand$ operation adds dimensions to a tensor by repeating its existing dimensions.
With this repetition, a large tensor can be represented efficiently in space and also give us a chance for further optimization (see \cref{section:expand}).
Such representation is implemented in tensor libraries like PyTorch \cite{Ansel_PyTorch_2_Faster_2024}.
For instance, the vector $[6\,7]$ can be $\expand$ed in a variety of ways:
\begin{align*}
    \expand([6\,7][x^{(2)}] \to x^{(2)}y^{(2)}) &= [\begin{smallmatrix}6&6\\7&7\end{smallmatrix}] &
    \expand([6\,7][x^{(2)}] \to x^{(2)}y^{(3)}) &= [\begin{smallmatrix}6&6&6\\7&7&7\end{smallmatrix}] \\
    \expand([6\,7][x^{(2)}] \to z^{(2)}x^{(2)}) &= [\begin{smallmatrix}6&7\\6&7\end{smallmatrix}] &
    \expand([6\,7][x^{(2)}] \to z^{(2)}x^{(2)}y^{(3)}) &= \bigl[[\begin{smallmatrix}6&6&6\\7&7&7\end{smallmatrix}]\,[\begin{smallmatrix}6&6&6\\7&7&7\end{smallmatrix}]\bigr]
\end{align*}
To take another example, the scalar~$1$ can be expanded to the vector $[1\,1] = \expand(1[]\to x^{(2)})$.

In~the general form $\expand(E[\vec{x}] \to \vec{z})$, the input index variables in~$\vec{x}$ must be distinct, the output index variables in~$\vec{z}$ must be distinct, and $\vec{x} \sqsubseteq \vec{z}$ (canonically, a~prefix of~$\vec{z}$, which is required by the optimization in \cref{opmt}).

\subsubsection{$\einsum$}
The $\einsum$ operation, or Einstein summation, generalizes scalar, vector, matrix, and elementwise multiplication, as can be seen in various tensor libraries \cite{Ansel_PyTorch_2_Faster_2024,tensorflow2015-whitepaper,harris2020array}.
These operations all define each output element as a sum of products of input elements.
In~order to specify which dimensions to sum over and which to multiply together, $\einsum$ uses index variables to relate the input tensors' dimensions to the output tensor's dimensions.
For instance, multiplying two matrices $E_1$ and~$E_2$ is denoted by $\einsum(E_1[xy],E_2[yz] \to xz)$.
Also, the binary and ternary multiplications in \labelcref{e:fgg-example} are expressed using $\einsum$:
\begin{align}
    \textrm{coin} &= \einsum(\tfrac{2}{3}[],[1\,0][x]\to x)
                   + \einsum(\tfrac{1}{3}[],[0\,1][x]\to x)\label{eq:coin}
\\
    \textrm{same} &= \einsum(\textrm{coin}[x], [\begin{smallmatrix}1&0\\0&1\end{smallmatrix}][xy], \textrm{coin}[y]\to)\label{eq:same}
\end{align}
Since \textrm{coin} is represented by a one-dimensional tensor, the two $\einsum$s on the right hand side of \cref{eq:coin} both have one output index variable $x$.
Similarly, since \textrm{same} is a scalar, the $\einsum$ in \cref{eq:same} doesn't have any output index variable.

In~the general form $\einsum(E_{1}[\vec{x}_{1}],\dotsc,E_{k}[\vec{x}_{k}] \to \vec{z})$, $\vec{z}$ can be empty, or the output index variables in~$\vec{z}$ must be distinct and must each appear at least once among $\vec{x}_{1},\dotsc,\vec{x}_{k}$, i.e., $\vec{z} \sqsubseteq \bigcup_{i=1}^k\vec{x}_i$.

\subsubsection{$\scatter$}
The $\scatter$ operation transforms a dense tensor to a sparse tensor by embedding the dense tensor in a larger one and padding the rest of the output with zero.
For instance, it~can form the diagonal matrix $[\begin{smallmatrix}6&0\\0&7\end{smallmatrix}]$ from the vector $[6\,7]$, it~can reshape a matrix into a vector, and it can enlarge a tensor along a dimension by surrounding it with zero padding.
Each dimension of the output tensor is described using an \emph{axis expression}~$e$. We explain axis expressions in \cref{section:fgg-scatter}, but for now, note that one base case is that an axis expression can be an index variable.
For instance, a~diagonal tensor is formed by repeating the same index variable qua axis expression multiple times:
\begin{equation}
    \scatter([6\,7][x] \to xx) = [\begin{smallmatrix}6&0\\0&7\end{smallmatrix}]
\end{equation}

In the general form $\scatter(E[\vec{x}] \to e\dotsm)$, the shape of the input tensor $E$ is the upper bounds of~$\vec{x}$, and the shape of the output sparse tensor is the upper bounds of $e\dotsm$.
We require syntactically that each index variable in~$\vec{x}$ appear at least once in~$e\dotsm$ somewhere.

\subsection{Denotational Semantics}
\label{section:fgg-denot}

For our purposes, a tensor is a multidimensional array of numbers, indexed by tuples of non-negative integers.
More specifically, each tensor has a \emph{shape}, which is a tuple of positive integers $(s_1,\dotsc,s_r)$, and the tuples $I=(i_1,\dotsc,i_r)$ that index the tensor satisfy $0\le i_1<s_1, \dotsc, 0\le i_r<s_r$.
The length $r$ of the shape is called the tensor's \emph{rank}; we say that the tensor has $r$ \emph{dimensions}, whose \emph{sizes} are $s_1,\dotsc,s_r$.
A~rank-0 tensor is a scalar; a~rank-1 tensor is a vector; a~rank-2 tensor is a matrix.
The element of~$T$ at the index~$I$ is $T[I]$.

An \emph{index function} is a function $\delta$ from some index variables to non-negative integers, such that $0 \le \delta(x^{(s)}) < s$ for each $x^{(s)}$ in the domain of~$\delta$.
Let $\vec{x} = x_1^{(s_1)} \dotsm x_r^{(s_r)}$ be a sequence of $r$ index variables, not necessarily distinct.
If~$\delta$ is an index function whose domain includes~$\vec{x}$, then we write $\delta(\vec{x})$ for the index tuple $\bigl( \delta(x_1), \dotsc, \delta(x_r) \bigr)$, which can be used to index a tensor of shape $(s_1,\dotsc,s_r)$.
Conversely, if~$I$ is an index tuple $(i_1,\dotsc,i_r)$, then we write $\{\vec{x}\mapsto I\}$ for the index function $\{ x_1\mapsto i_1, \dotsc, x_r\mapsto i_r \}$; here we do require the index variables in~$\vec{x}$ to be distinct.
We also define the set $\dom{\vec{x}}$ to consist of all index functions whose domain is~$\vec{x}$, and the set $\tups{\vec{x}}$ to consist of all index tuples $(i_1,\dotsc,i_r)$ that satisfy $0\le i_1<s_1, \dotsc, 0\le i_r<s_r$.

The denotation of a tensor expression~$E$ depends on the denotations of the tensor variables in~$E$.
So let the environment $\gamma$ be a function from tensor variables to tensors, and define the denotation $\llbracket E \rrbracket_\gamma$ as follows:
{
\allowdisplaybreaks
\begin{alignat}{2}
    &\llbracket t \rrbracket_\gamma [I] &&=\gamma(t) [I]
\\
\label{e:fgg-scalar}
    &\llbracket \alpha \rrbracket_\gamma [()] &&=\alpha
\\
\label{e:fgg-add}
    &\llbracket E_1 + E_2 \rrbracket_\gamma [I] &&=\llbracket E_1 \rrbracket_\gamma [I] + \llbracket E_2 \rrbracket_\gamma [I]
\\
    &\llbracket 0^{(s,\dotsc)} \rrbracket_\gamma [I] &&=0
\\
\label{e:fgg-expand}
    &\llbracket \expand(E[\vec{x}] \to \vec{z}) \rrbracket_\gamma [I]
    &&=\llbracket E \rrbracket_\gamma [\{\vec{z}\mapsto I\}(\vec{x})]
\\
\label{e:fgg-einsum}
    &\llbracket \einsum(E_1[\vec{x}_1],\dotsc,E_k[\vec{x}_k] \to \vec{z}) \rrbracket_\gamma [I]
    &&=\sum\nolimits_{\begin{subarray}{l}\delta\in\dom{\bigcup_{i=1}^k\vec{x}_i} \\ \delta(\vec{z})=I\end{subarray}} \prod\nolimits_{i=1}^k \llbracket E_i \rrbracket_\gamma [\delta(\vec{x}_i)]
\\
\label{e:fgg-scatter}
    &\llbracket \scatter(E[\vec{x}] \to e\dotsm) \rrbracket_\gamma [I]
    &&=\sum\nolimits_{\begin{subarray}{l}\delta\in\dom{\vec{x}} \\ \delta(e\dotsm) = I\end{subarray}} \llbracket E \rrbracket_\gamma [\delta(\vec{x})]
\end{alignat}
}

The first four cases are simple.
\Cref{e:fgg-scalar} says that a literal scalar has rank~$0$.
\Cref{e:fgg-add} requires two tensors to have the same shape in order to add them.

\Cref{e:fgg-expand} says that, to~look up an element of an $\expand$ed tensor, we~just look up the corresponding element of the original tensor, by~removing from the index tuple the new dimensions introduced by $\expand$.
For instance, the element of $\llbracket \expand(E[x] \to xy) \rrbracket_\gamma$ at $(i,j)$ is defined to be the element of $\llbracket E \rrbracket_\gamma$ at~$(i)$.

\Cref{e:fgg-einsum} defines the denotation of an $\einsum$ expression by iterating over all index tuples and summing out those index variables that do not appear in the output~$\vec{z}$. (Recall that our syntax requires $\vec{z}$ to be a subset of $\bigcup_{i=1}^k\vec{x}_i$.)
For instance, multiplying two matrices $E_1$ and~$E_2$ is denoted by $\einsum(E_1[xy],E_2[yz] \to xz)$, which by \labelcref{e:fgg-einsum} amounts to
\begin{equation}
    \llbracket\einsum(E_1[xy],E_2[yz] \to xz)\rrbracket_\gamma[(i,k)] = \sum\nolimits_j \llbracket E_1\rrbracket_\gamma[(i,j)] \llbracket E_2\rrbracket_\gamma[(j,k)].
\end{equation}
In the special case where $\vec{z} = \bigcup_{i=1}^k\vec{x}_i$, we say that the $\einsum$ is \emph{summation-free}:
\begin{equation}
\label{e:fgg-einsum-summation-free}
    \llbracket \einsum(E_1[\vec{x}_1],\dotsc,E_k[\vec{x}_k] \to \vec{z}) \rrbracket_\gamma [I]
    = \prod\nolimits_{i=1}^k \llbracket E_i \rrbracket_\gamma [\{\vec{z} \mapsto I\}(\vec{x}_i)]
\end{equation}
The right hand side of~\labelcref{e:fgg-einsum-summation-free} does not quite reduce to the right hand side of~\labelcref{e:fgg-expand} when $k=1$, because $\expand$ allows $\vec{z}$ to contain index variables not present in~$\vec{x}.$
This denotational semantics of $\einsum$ is similar to the one given by \citet{wenigSyntaxSemanticsEinsum2025}.

\Cref{e:fgg-scatter}, for $\scatter$, involves axis expressions, whose explanation we defer to \cref{section:fgg-scatter} below.
For now, we~promise that the summation on the right hand side iterates over at most one term; in~other words, the $\delta\in\dom{\vec{x}}$ such that $\delta(e\dotsm) = I$ is unique.
Consider the base case where $e\dotsm$ is a sequence entirely comprised of index variables.
We~require that $\vec{x}$ and~$e\dotsm$ contain the exact same set of index variables, and $\vec{x}$ contain no duplicates.
For instance, it~could be that $\vec{x}=x$ and $e\dotsm=xx$.
Then, the index function $\delta\in\dom{x}$ satisfies $\delta(e\dotsm) = \delta(xx) = I$ just in case $\delta=\{x\mapsto i\}$ and $I=(i,i)$ for some index~$i$.
Hence \labelcref{e:fgg-scatter} forms a diagonal matrix from a vector:
\begin{equation}
    \llbracket \scatter(E[x] \to xx) \rrbracket_\gamma [(i,j)] =
    \begin{cases}
        \llbracket E \rrbracket_\gamma [(i)] & \text{if $i=j$} \\
        0 & \text{otherwise}
    \end{cases}
\end{equation}

\subsection{Representing Sparsity}
\label{section:fgg-scatter}

Besides expansion and Einstein summation, our compiler makes heavy use of tensors with sparsity patterns such as diagonal matrices, zero-padded dimensions, and Kronecker products, in order to handle sum and product types.
To~represent such tensors compactly and operate on them efficiently, \citet{chiangImplementation2025} introduce \emph{axis expressions}.

Intuitively, an~axis expression denotes an injective function from some indices (into the input tensor of $\scatter$) to an index (into the output tensor of $\scatter$).
For instance, the tensor expression $\scatter\bigl(E[x^{(s)}] \to (s_1 \oplus x^{(s)} \oplus s_2)\bigr)$ means to pad a vector~$E$ with $s_1$ zeros at the beginning and $s_2$ zeros at the end, because the axis expression $s_1 \oplus x^{(s)} \oplus s_2$ denotes the injective function from the index function $\{x\mapsto i\}$ to the index $s_1+i$, where $i<s$.
And the tensor expression $\scatter\bigl(E[x^{(s_1)} y^{(s_2)}] \to (x^{(s_1)} \otimes y^{(s_2)})\bigr)$ means to reshape a matrix~$E$ into a vector, because the axis expression $x^{(s_1)} \otimes y^{(s_2)}$ denotes the injective function from the index function $\{x\mapsto i,y\mapsto j\}$ to the index $i\cdot s_2 + j$, where $i<s_1$ and $j<s_2$.

The general grammar for axis expressions~$e$ is shown in \cref{fig:fgg-syntax}.
Each axis expression~$e$ has an \emph{upper bound}~$|e|$, which generalizes the upper bound of index variables:
\begin{align}
    |x^{(s)}| &= s &
    |()| &= 1 &
    |s_1 \oplus e \oplus s_2| &= s_1 + |e| + s_2 &
    |e_1 \otimes e_2| &= |e_1| \cdot |e_2|
\intertext{An axis expression $e$ denotes an integer $\llbracket e \rrbracket_\delta$ between $0$ and $|e|-1$, depending on the index function~$\delta$:}
    \llbracket x \rrbracket_\delta &= \delta(x) &
    \llbracket () \rrbracket_\delta &= 0 &
    \llbracket s_1 \oplus e \oplus s_2 \rrbracket_\delta &= s_1 + \llbracket e \rrbracket_\delta &
    \llbracket e_1 \otimes e_2 \rrbracket_\delta &= \llbracket e_1 \rrbracket_\delta \cdot |e_2| + \llbracket e_2 \rrbracket_\delta
\end{align}
For a sequence of axis expressions $e\dotsm$, we write $\delta(e\dotsm) = \llbracket e\rrbracket_\delta \dotsm$.

Recall that in $\scatter(E[\vec{x}] \to e\dotsm)$, each index variable in~$\vec{x}$ appears at least once in~$e\dotsm$ somewhere, so that the function from $\delta\in\dom{\vec{x}}$ to $\delta(e\dotsm)$ is injective.
That's why the summation on the right hand side of \cref{e:fgg-scatter} iterates over at most one term.

\section{Compilation}\label{comp}

We describe our compiler from ISO to FGGs, then prove that it preserves semantics.
One feature of our approach is that the compilation is relatively straightforward to implement (about 1000 lines of Haskell).

\subsection{Compiling ISO to FGGs}

\begin{figure}
\centering
\begin{equation*}
  \begin{aligned}
    &\data{comp}{}{} &::~ & \data{Term}{iso}{} \to \data{Tensor}{}{} & & \\
    &\comp{\kw{()}} &={} & \expand(1[]\to x_{\mathbbold{1}})\\
    &\comp{x}       &={} & \scatter(\expand(1[]\to x_{a})[x_{a}] \to x_{a}x_{a})\\
    &\comp{\kw{inj}_{l}~t} &={} & \einsum(\comp{t}[\comp{\Delta}x_{a}],\\
    &&&\hphantom{\einsum{}(}\scatter(\expand(1[] \to x_{a})[x_{a}] \to\\
    &&&\hphantom{\einsum{}(\scatter{}(}x_{a}(0\oplus{x_{a}}\oplus\comp{b}))[x_{a}z_{a\oplus{b}}] \to\\
    &&&\hphantom{\einsum{}(}{\comp{\Delta}}^{}{z_{a\oplus{b}}}^{})\\
    &&&\text{if}~\typeof{\kw{inj}_{l}~t}{a \oplus b} \\
    &\comp{\kw{inj}_{r}~t} &={} & \einsum(\comp{t}[\comp{\Delta}y_{b}],\\
    &&&\hphantom{\einsum{}(}\scatter(\expand(1[] \to y_{b})[y_{b}] \to\\
    &&&\hphantom{\einsum{}(\scatter{}(}y_{b}(\comp{a}\oplus{y_{b}}\oplus0))[y_{b}z_{a\oplus{b}}] \to\\
    &&&\hphantom{\einsum{}(}{\comp{\Delta}}^{}{z_{a\oplus{b}}}^{})\\
    &&& \text{if}~\typeof{\kw{inj}_{r}~t}{a \oplus b} \\
    &\comp{\langle{t_{1}, t_{2}}\rangle}
                     &={} & \einsum(\comp{t_{1}}[\comp{\Delta_{1}}x_{a}],\comp{t_{2}}[\comp{\Delta_{2}}y_{b}],\\
    &&&\hphantom{\einsum{}(}\scatter(\expand(1[] \to x_{a}y_{b})[x_{a}y_{b}] \to\\
    &&&\hphantom{\einsum{}(\scatter{}(} x_{a}y_{b}(x_{a} \otimes y_{b}))[x_{a}y_{b}z_{a\otimes{b}}] \to\\
    &&&\hphantom{\einsum{}(}\comp{\Delta_{1}}\comp{\Delta_{2}}z_{a\otimes{b}})\\
    &&&~\text{if}~\typeof{\langle{t_{1}, t_{2}}\rangle}{a \otimes b} \\
    &\comp{\omega~t}
                     &={} & \einsum(\comp{t}[\comp{\Delta}x_{a}],\comp{\omega}[x_{a}y_{b}] \to \comp{\Delta}y_{b})\\
    &\comp{\kw{let}~x~\kw{=}~t_{1}~\kw{in}~t_{2}}
                     &={} & \comp{t_{2}[t_{1}/x]}\\
    &\comp{\kw{let}~p~\kw{=}~t_{1}~\kw{in}~t_{2}}
                     &={} & \einsum(\comp{t_{1}}[\comp{\Delta_{1}}x_{a}],\comppat{p}[\comp{\Delta_{2}'}x_{a}],\\
    &&&\hphantom{\einsum{}(}\comp{t_{2}}[\comp{\Delta_{2}}\comp{\Delta_{2}'}y_{b}] \to\\
    &&&\hphantom{\einsum{}(}\comp{\Delta_{1}}\comp{\Delta_{2}}y_{b}) \\
    &\comp{t_{1} + t_{2}}
                    &={} & \comp{t_{1}} + \comp{t_{2}} \\
    &\comp{\alpha{t}}
                     &={} & \einsum(\alpha[],\comp{t}[\comp{{\Delta}}x_{\tau}] \to \comp{\Delta}x_{\tau})\\
    &\\
    &\data{comp}{}{} &::~ & \data{Iso}{iso}{} \to \data{Tensor}{}{} \\
    &\comp{
      \left\{
        \case v_{1} \leftrightarrow e_{1} \cdots \case v_{n} \leftrightarrow e_{n}
      \right\}} &={} &\comp{v_{1} \leftrightarrow e_{1}} + \cdots + \comp{v_{n} \leftrightarrow e_{n}}\\
    &\text{where}\\
    &\comp{v \leftrightarrow e}
                     &={} & \einsum(\compv{v}[\comp{\Delta}x_{a}], \compext{e}[\comp{\Delta}y_{b}] \to x_{a}y_{b})\\
    &\\
    &\data{comp}{}{} &::~ &\data{Prog}{iso}{} \to \data{Prog}{fgg}{} \\
    &\comp{\omega} &={} & (\textrm{main} = \comp{\omega})\\
    &\comp{t} &={} & (\textrm{main} = \comp{t})\\
  \end{aligned}
\end{equation*}
\caption{The compiler from ISO terms and isos to FGGs}
\label{fig:compiler}
\end{figure}
\begin{figure}
\begin{equation*}
  \begin{aligned}
    &\rw{Hadamard*} ::
      \mathbbold{3} \oplus \mathbbold{3} \leftrightarrow \mathbbold{3} \oplus \mathbbold{3}\\
    &\left\{
      \begin{aligned}
        &\case \kw{inj}_{l}~x \leftrightarrow
          \tfrac{1}{\sqrt{2}}(\kw{inj}_{l}~x) + \tfrac{1}{\sqrt{2}}(\kw{inj}_{r}~x);\\
        &\case \kw{inj}_{r}~x \leftrightarrow
          \tfrac{1}{\sqrt{2}}(\kw{inj}_{l}~x) - \tfrac{1}{\sqrt{2}}(\kw{inj}_{r}~x)
      \end{aligned}
      \right\}\\
    &\rw{Hadamard*}~(\kw{inj}_{l}~(\kw{inj}_{r}()))
  \end{aligned}
\end{equation*}
\newcounter{oldEqCounter}
\setcounter{oldEqCounter}{\value{equation}}
{
\renewcommand{\theequation}{\alph{equation}}
\renewcommand{\theHequation}{\alph{equation}}
\setcounter{equation}{0}
\begin{align}
  \leftbox{em:had-app}{$\rw{Hadamard*}~\dot{\mathbf{2}}$}
  \label{eg:fgg-had-app}
  \textrm{main} &=
  \einsum(
  E_{\dot{\mathbf{2}}}[x_{\mathbbold{6}}^{(6)}],
  E_{H}[x_{\mathbbold{6}}^{(6)}y_{\mathbbold{6}}^{(6)}]
    \to
    y_{\mathbbold{6}}^{(6)})
  \\
  \leftbox{em:had}{$\rw{Hadamard*}$}
  E_{H} &=
  \einsum(E_{\FF}[w_{\mathbbold{3}}^{(3)}x_{\mathbbold{6}}^{(6)}],
          E_{+}[w_{\mathbbold{3}}^{(3)}y_{\mathbbold{6}}^{(6)}] \to
          x_{\mathbbold{6}}^{(6)}y_{\mathbbold{6}}^{(6)}) \notag\\[-\jot]&\qquad{}+
  \einsum(E_{\TT}[w_{\mathbbold{3}}^{(3)}x_{\mathbbold{6}}^{(6)}],
          E_{-}[w_{\mathbbold{3}}^{(3)}y_{\mathbbold{6}}^{(6)}] \to
          x_{\mathbbold{6}}^{(6)}y_{\mathbbold{6}}^{(6)})
  \label{eg:fgg-had}
  \\
  \label{eg:fgg-plus}
  \leftbox{em:plus}{\scriptsize{$\tfrac{1}{\sqrt{2}}\kw{inj}_{l}~x + \tfrac{1}{\sqrt{2}}\kw{inj}_{r}~x$}}
  E_{+}
  &=
  E_{\frac{1}{\sqrt{2}}\FF} + E_{\frac{1}{\sqrt{2}}\TT}
  \\
  \label{eg:fgg-minus}
  \leftbox{em:minus}{\scriptsize{$\tfrac{1}{\sqrt{2}}\kw{inj}_{l}~x - \tfrac{1}{\sqrt{2}}\kw{inj}_{r}~x$}}
  E_{-}
  &=
  E_{\frac{1}{\sqrt{2}}\FF} + E_{-\frac{1}{\sqrt{2}}\TT}
  \\
  \label{eg:fgg-sqrt-ff}
  \leftbox{em:sqrt-ff}{$\tfrac{1}{\sqrt{2}}\kw{inj}_{l}~x$}
  E_{\frac{1}{\sqrt{2}}\FF}
  &=
  \einsum(
  \tfrac{1}{\sqrt{2}}[],
  E_{\FF}[w_{\mathbbold{3}}^{(3)}z_{\mathbbold{6}}^{(6)}]
  \to
  w_{\mathbbold{3}}^{(3)}z_{\mathbbold{6}}^{(6)})
  \\
  \label{eg:fgg-sqrt-tt}
  \leftbox{em:sqrt-tt}{$\tfrac{1}{\sqrt{2}}\kw{inj}_{r}~x$}
  E_{\frac{1}{\sqrt{2}}\TT}
  &=
  \einsum(
  \tfrac{1}{\sqrt{2}}[],
  E_{\TT}[w_{\mathbbold{3}}^{(3)}z_{\mathbbold{6}}^{(6)}]
  \to
  w_{\mathbbold{3}}^{(3)}z_{\mathbbold{6}}^{(6)})
  \\
  \label{eg:fgg-msqrt-tt}
  \leftbox{em:msqrt-tt}{$-\tfrac{1}{\sqrt{2}}\kw{inj}_{r}~x$}
  E_{-\frac{1}{\sqrt{2}}\TT}
  &=
  \einsum(
  -\tfrac{1}{\sqrt{2}}[],
  E_{\TT}[w_{\mathbbold{3}}^{(3)}z_{\mathbbold{6}}^{(6)}]
  \to
  w_{\mathbbold{3}}^{(3)}z_{\mathbbold{6}}^{(6)})
  \\
  \leftbox{em:ff}{$\kw{inj}_{l}~x$}
  \label{eg:fgg-ff}
  E_{\FF} &=
  \einsum(
  E_{x}[w_{\mathbbold{3}}^{(3)}x_{\mathbbold{3}}^{(3)}],
  E_{l}[x_{\mathbbold{3}}^{(3)}z_{\mathbbold{6}}^{(6)}]
  \to
  w_{\mathbbold{3}}^{(3)}z_{\mathbbold{6}}^{(6)})
  \\
  \leftbox{em:tt}{$\kw{inj}_{r}~x$}
  \label{eg:fgg-tt}
  E_{\TT} &=
  \einsum(
  E_{x}[w_{\mathbbold{3}}^{(3)}y_{\mathbbold{3}}^{(3)}],
  E_{r}[y_{\mathbbold{3}}^{(3)}z_{\mathbbold{6}}^{(6)}]
  \to
  w_{\mathbbold{3}}^{(3)}z_{\mathbbold{6}}^{(6)})
  \\
  \label{eg:fgg-var}
  \leftbox{em:x}{$x$}
  E_{x}
  &=
    \scatter(\expand(1[]\to x_{\mathbbold{3}}^{(3)})[x_{\mathbbold{3}}^{(3)}]\to
    x_{\mathbbold{3}}^{(3)}x_{\mathbbold{3}}^{(3)})
  \\
  \label{eg:fgg-injl6}
  \leftbox{em:injl6}{$\kw{inj}_{l}~(\kw{inj}_{r}~())$}
  E_{\dot{\mathbf{2}}}
  &=
  \einsum(
  E_{\kw{inj}_{r}()}[x_{\mathbbold{3}}^{(3)}],
  E_{l}[x_{\mathbbold{3}}^{(3)}z_{\mathbbold{6}}^{(6)}]
  \to
  z_{\mathbbold{6}}^{(6)})
  \\
  \label{eg:fgg-injr6}
  \leftbox{em:injr3}{$\kw{inj}_{r}~()$}
  E_{\kw{inj}_{r}()}
  &=
  \einsum(
  E_{()}[x_{\mathbbold{1}}^{(1)}],
  E_{r'}[x_{\mathbbold{1}}^{(1)}z_{\mathbbold{3}}^{(3)}]
  \to
  z_{\mathbbold{3}}^{(3)})
  \\
  \label{eg:fgg-unit}
  \leftbox{em:unit}{$()$}
  E_{()}
  &=
  \expand(1[]\to x_{\mathbbold{1}}^{(1)})
  \\
  \label{eg:fgg-left}
  \tikzmark{em:left}%
  E_{l}
  &=
  \scatter(
  \expand(1[] \to x_{\mathbbold{3}}^{(3)})[x_{\mathbbold{3}}^{(3)}]
  \to
  x_{\mathbbold{3}}^{(3)}(0\oplus{x_{\mathbbold{3}}^{(3)}}\oplus{3}))
  \\
  \label{eg:fgg-right}
  \tikzmark{em:right}%
  E_{r}
  &=
  \scatter(
  \expand(1[] \to x_{\mathbbold{3}}^{(3)})[x_{\mathbbold{3}}^{(3)}]
  \to
  x_{\mathbbold{3}}^{(3)}(3\oplus{x_{\mathbbold{3}}^{(3)}}\oplus{0}))
  \\
  \label{eg:fgg-right-1}
  \tikzmark{em:fgg-right-1}
  E_{r'}
  &=
  \scatter(
  \expand(1[] \to x_{\mathbbold{1}}^{(1)})[x_{\mathbbold{1}}^{(1)}]
  \to
  x_{\mathbbold{1}}^{(1)}(2\oplus{x_{\mathbbold{1}}^{(1)}}\oplus{0}))
\end{align}
}
\setcounter{equation}{\value{oldEqCounter}}
\[
  \renewcommand{\arraystretch}{2.3}
  \begin{array}{llll}
    \labelcref{eg:fgg-had-app}~
    \frac{1}{\sqrt{2}}\left[
    \begin{smallmatrix}
      0\\
      0\\
      1\\
      0\\
      0\\
      1
    \end{smallmatrix}\right]
    & \labelcref{eg:fgg-had}~
      \frac{1}{\sqrt{2}}\left[
      \begin{smallmatrix}
        1&0&0&1&0&0\\
        0&1&0&0&1&0\\
        0&0&1&0&0&1\\
        1&0&0&-1&0&0\\
        0&1&0&0&-1&0\\
        0&0&1&0&0&-1\\
      \end{smallmatrix}\right]
    & \labelcref{eg:fgg-plus}~
      \tfrac{1}{\sqrt{2}}\left[
      \begin{smallmatrix}
        1&0&0&1&0&0\\
        0&1&0&0&1&0\\
        0&0&1&0&0&1\\
      \end{smallmatrix}\right]
    & \labelcref{eg:fgg-minus}~
      \tfrac{1}{\sqrt{2}}\left[
      \begin{smallmatrix}
        1&0&0&-1&0&0\\
        0&1&0&0&-1&0\\
        0&0&1&0&0&-1\\
      \end{smallmatrix}\right]\\
    \labelcref{eg:fgg-sqrt-ff}~
    \frac{1}{\sqrt{2}}\left[
    \begin{smallmatrix}
      1&0&0&0&0&0\\
      0&1&0&0&0&0\\
      0&0&1&0&0&0
    \end{smallmatrix}\right]
    & \labelcref{eg:fgg-sqrt-tt}~
      \tfrac{1}{\sqrt{2}}\left[
      \begin{smallmatrix}
        0&0&0&1&0&0\\
        0&0&0&0&1&0\\
        0&0&0&0&0&1\\
      \end{smallmatrix}\right]
    & \labelcref{eg:fgg-msqrt-tt}~
      \tfrac{1}{\sqrt{2}}\left[
      \begin{smallmatrix}
        0&0&0&-1&0&0\\
        0&0&0&0&-1&0\\
        0&0&0&0&0&-1
      \end{smallmatrix}\right]
    & \labelcref{eg:fgg-ff}~
      \left[
      \begin{smallmatrix}
        1&0&0&0&0&0\\
        0&1&0&0&0&0\\
        0&0&1&0&0&0
      \end{smallmatrix}\right]\\
    \labelcref{eg:fgg-tt}~
    \left[
    \begin{smallmatrix}
      0&0&0&1&0&0\\
      0&0&0&0&1&0\\
      0&0&0&0&0&1
    \end{smallmatrix}\right]
    & \labelcref{eg:fgg-var}~
      \left[
      \begin{smallmatrix}
        1&0&0\\
        0&1&0\\
        0&0&1\\
      \end{smallmatrix}\right]
    & \labelcref{eg:fgg-injl6}~
      \left[
      \begin{smallmatrix}
        0\\
        0\\
        1\\
        0\\
        0\\
        0
      \end{smallmatrix}\right]
    & \labelcref{eg:fgg-injr6}~
      \left[
      \begin{smallmatrix}
        0\\
        0\\
        1
      \end{smallmatrix}\right]\\
    \labelcref{eg:fgg-unit}~
    \left[
    \begin{smallmatrix}
      1
    \end{smallmatrix}
    \right]
    & \labelcref{eg:fgg-left}~
      \left[
      \begin{smallmatrix}
        1&0&0&0&0&0\\
        0&1&0&0&0&0\\
        0&0&1&0&0&0
      \end{smallmatrix}\right]
    & \labelcref{eg:fgg-right}~
      \left[
      \begin{smallmatrix}
        0&0&0&1&0&0\\
        0&0&0&0&1&0\\
        0&0&0&0&0&1
      \end{smallmatrix}\right]
    & \labelcref{eg:fgg-right-1}~
      \tfrac{1}{\sqrt{2}}
      \left[
      \begin{smallmatrix}
        0 & 0 & 1
      \end{smallmatrix}
      \right]
  \end{array}
\]
\vspace{-1em}
\caption{The ISO program in \cref{eg:had*} with $a = (\mathbbold{1}\oplus\mathbbold{1})\oplus{\mathbbold{1}}$ and an application $\rw{Hadamard*}~(\kw{inj}_{l}~(\kw{inj}_{r}()))$ can be compiled to the tensor equations above.
  We use $\mathbbold{3}$ to represent the type $(\mathbbold{1}\oplus\mathbbold{1})\oplus{\mathbbold{1}}$, $\mathbbold{6}$ to represent the type $\mathbbold{3}\oplus\mathbbold{3}$, and $\dot{\mathbf{2}}$ to represent the value $\kw{inj}_{l}~(\kw{inj}_{r}())$ of type $\mathbbold{6}$.
  These equations are labeled with the corresponding subexpressions from \cref{eg:had*} on the left.
  The corresponding tensors are listed below the equations.
  \cref{eg:fgg-had-app,eg:fgg-injl6,eg:fgg-injr6,eg:fgg-unit} represent rank-1 tensors; all other equations represent rank-2 tensors.}
\label{fig:had-compile}
\end{figure}

The key idea of our compiler is based on the denotational semantics of ISO.
In \cref{sec:iso-denot}, we define the denotational semantics of ISO as a function from environments and semantic values to weights.
For any given well-typed expression, if we iterate over all possible environments and semantic values, and tabulate the corresponding weights, we get a tensor that represents this function.
For example, the term $\TyIsoTerm{x:\mathbbold{3}}{\kw{inj}_{l}~x}{\mathbbold{6}}$ is typed in the context $x:\mathbbold{3}$, where $\mathbbold{3}$ is a shorthand for the type $(\mathbbold{1} \oplus \mathbbold{1}) \oplus \mathbbold{1}$, and $\mathbbold{6}$ is a shorthand for the type $\mathbbold{3} \oplus \mathbbold{3}$.
An~environment corresponding to $x:\mathbbold{3}$ maps $x$ to $\kw{inj}_{l}~(\kw{inj}_{l}~())$, $\kw{inj}_{l}~(\kw{inj}_{r}~())$, or $\kw{inj}_{r}~()$.
And there are six semantic values of type $\mathbbold{6}$.
Therefore, tabulating the possible weights of this term gives the $3 \times 6$ tensor %
$\left[
  \begin{smallmatrix}
    1&0&0&0&0&0\\
    0&1&0&0&0&0\\
    0&0&1&0&0&0
  \end{smallmatrix}\right]$.
Based on this idea, we define a translation from ISO expressions to FGG tensor expressions.
The compiler for terms and isos is shown in \cref{fig:compiler}.
The compiler for patterns, pure values, and extended values is shown in Appendix~\ref{section:comp-iso-more}.
A complete example that is based on \cref{eg:had*} and extended from $\TyIsoTerm{x:\mathbbold{3}}{\kw{inj}_{l}~x}{\mathbbold{6}}$ is shown in \cref{fig:had-compile}.

\subsubsection{Compilation of types and typing contexts}\label{sec:comp-type}

From \cref{sec:iso-denot}, we know that every ISO value type corresponds to a finite set of semantic values.
Since a semantic value can be thought of as an index, every type is compiled to a natural number---the size of the set---that indicates the upper bound of an index variable.
The compiler for ISO value types is defined as follows.
\begin{alignat}{2}
    &\comp{\mathbbold{1}} &&= 1\\
    &\comp{a \oplus b} &&= \comp{a} + \comp{b}\\
    &\comp{a \otimes b} &&= \comp{a} \cdot \comp{b}
\end{alignat}
Following the definition above, we can compile a typing context to an index variable sequence.
\begin{alignat}{2}
    &\comp{\varnothing} &&= ()\\
    &\comp{\Delta,x:a} &&= \comp{\Delta} x_{a}^{(\comp{a})}
\end{alignat}

\subsubsection{Compilation of terms}

The term compiler $\data{comp}{}{}$ in \cref{fig:compiler} takes a well-typed term $\TyIsoTerm{\Delta}{t}{a}$ and returns an FGG tensor expression of rank $|\Delta| + 1$.
We write $t$ instead of the whole typing judgment for brevity.
For a well-typed term $t$, its denotation takes an environment corresponding to its typing context $\Delta$ and a semantic value of type $a$.
Therefore, the rank of the compiled tensor is $|\Delta| + 1$.
Each variable $x:\tau$ in $\Delta$ has a corresponding index variable $x_{\tau}^{(\comp{\tau})}$.
The type $a$ corresponds to another fresh index variable $x_{a}^{(\comp{a})}$.
We omit the upper bound $^{(s)}$ from an index variable~$x^{(s)}$.
Since pure values, extended values, and patterns are all subsets of terms, we relegate their compilation to Appendix~\ref{section:comp-iso-more}.
We introduce $\data{comp}{}{}$ case by case below.

The unit term \kw{()} is compiled to an expanded vector $[1]$, as shown in \cref{eg:fgg-unit} in \cref{fig:had-compile}.

A variable term $x$ is typed as $\TyIsoTerm{x:a}{x}{a}$.
As shown in \cref{sec:iso-denot}, the denotational semantics of such a term takes an environment $\{(x,\val{}')\}$ and a semantic value $\val{}$ and returns 1 only when $\val{}' = \val{}$.
Since both $\val{}$ and $\val{}'$ have type $a$, we should compile $x$ by making an identity matrix.
Therefore, we start with the scalar $1$ and extend it to a vector of size $\comp{a}$ by $\expand(1[] \to x_{a})$.
Then we further extend it to an identity matrix by $\scatter(\expand(1[]\to x_{a})[x_{a}] \to x_{a}x_{a})$.
\Cref{eg:fgg-var} in \cref{fig:had-compile} shows an example.

The sum terms $\kw{inj}_{l}~t$ and $\kw{inj}_{r}~t$ are compiled in the same way, as shown in \cref{eg:fgg-ff,eg:fgg-tt} in \cref{fig:had-compile}.
For $\TyIsoTerm{x_{1}:\tau_{1}, \cdots, x_{n}:\tau_{n}}{\kw{inj}_{l}~t}{a \oplus b}$, we first compile $t$ to a tensor of the shape $(\comp{\tau_{1}},\cdots,\comp{\tau_{n}},\comp{a})$.
This tensor is then passed as an argument to $\einsum$.
The second argument to $\einsum$ is a sparse rank-2 tensor that has value 1 where semantic values from types $a$ and $a \oplus b$ match.
For instance, if $a$ is $\mathbbold{1}$ and $b$ is $\mathbbold{1} \otimes \mathbbold{1}$, for semantic values $()$ and $\kw{inj}_{l}~()$, this tensor has value 1; for $()$ and $\kw{inj}_{r}~((),())$, it has value 0.
The result of $\einsum$ is a tensor of the shape $(\comp{\tau_{1}},\cdots,\comp{\tau_{n}},\comp{a}+\comp{b})$.

The product term $\TyIsoTerm{\Delta_{1},\Delta_{2}}{\langle{t_{1},t_{2}}\rangle}{a\otimes{b}}$ can be compiled in a similar way.
The difference is we need to recursively compile two terms.
Moreover, the third argument to $\einsum$ is a sparse rank-3 tensor that has 1 where semantic values from types $a$, $b$, and $a\otimes{b}$ match.

The application term $\omega~t$ is compiled to an $\einsum$ whose two arguments are the results from recursively compiling the operator and the operand, as shown in \cref{eg:fgg-had-app} in \cref{fig:had-compile}.
See \cref{compilation-of-isos} for how to compile an iso.

The $\kw{let}$ term is processed depending on what the left hand side pattern is.
If the pattern is a single variable $x$, we first substitute the right hand side term $t_{1}$ for $x$ in the body $t_{2}$, then we compile the substituted term $t_{2}[t_{1}/x]$.
Such substitutions turn out to be beneficial for the FGG solver to detect sparsity in the quantum simulation.
See \cref{bench} for an analysis of this optimization.
Otherwise, the $\kw{let}$ expression is processed in a similar way to application.
The difference is there are three subterms in a $\kw{let}$.
Therefore, the compiled $\einsum$s have three arguments.

The linear combination term $t_{1} + t_{2}$ is compiled to a tensor sum, as shown in \cref{eg:fgg-plus,eg:fgg-minus}.

The weighted term $\alpha{t}$ is compiled to an $\einsum$ that multiplies a tensor by a scalar, as shown in \cref{eg:fgg-sqrt-ff,eg:fgg-sqrt-tt,eg:fgg-msqrt-tt}.

\subsubsection{Compilation of isos}
\label{compilation-of-isos}

To compile an iso, we compile each clause to a tensor and sum them together.
This compilation is therefore similar to a linear combination term.
Each clause is an alternative outcome and compiled to an $\einsum$ with two arguments, which correspond to the pure value and the extended value respectively.
Since the pure value and the extended value are typed in the same typing context, they share the indices corresponding to the same environment.
The other index for them comes from their types.
The result of the $\einsum$ is a rank-2 tensor with indices from the types of pure values and extended values.
\Cref{eg:fgg-had} in \cref{fig:had-compile} shows an example of compiling an iso.

\subsubsection{Compilation of programs}

We compile an ISO program to an FGG equation.
We generate a new tensor variable as the left hand side and compile the program to a tensor as the right hand side.

As defined in \cref{fig:compiler}, the compiler turns repeated ISO subexpressions into repeated FGG subexpressions. Our implementation performs common subexpression elimination by introducing a distinct tensor variable for each unique ISO subexpression.

\subsection{Correctness of the Compilation}

To state the correctness of the compilation, we first define the compilation of semantic values and ISO environments.

For each type~$a$, we define a bijection $\compsv{a}{\cdot}$ between the semantic values in $\denot{a}$ and the natural numbers in $\{0,\dotsc,\comp{a}-1\}$:
{
\allowdisplaybreaks
\begin{alignat}{2}
    &\compsv{\mathbbold{1}}{()} &&= 0 \\
    &\compsv{a \oplus b}{\kw{inj}_{l}~\val{}} &&= \compsv{a}{\val{}} \\
    &\compsv{a \oplus b}{\kw{inj}_{r}~\val{}} &&= \comp{a} + \compsv{b}{\val{}} \\
    &\compsv{a \otimes b}{\langle\val{1},\val{2}\rangle} &&= \compsv{a}{\val{1}} \cdot \comp{b} + \compsv{b}{\val{2}}
\end{alignat}
}%
This definition orders semantic values lexicographically, but any systematic ordering will work, as~long as the $\kw{inj}_{l}$, $\kw{inj}_{r}$, and $\langle{\_,\_}\rangle$ cases in \cref{fig:compiler} are adjusted accordingly.

Based on \fcompsv, for each typing context~$\Delta$, we define a bijection $\compenv{\Delta,\cdot}$ between ISO environments in $\denot{\Delta}$ and index tuples in $\tups{\comp{\Delta}}$:
\begin{alignat}{2}
    &\compenv{\varnothing,\varnothing} &&= ()\\
    &\compenv{(\Delta,x:a),(\delta\cup\{(x,\val{})\})} &&= \compenv{\Delta,\delta},\compsv{a}{\val{}}
\end{alignat}

Then we can state the following theorem.

\begin{theorem}\label{thm:wty-denot}
  The compilation in \cref{fig:compiler} preserves the denotational semantics of well-typed ISO programs:
  \begin{itemize}
  \item If $\,\TyIsoTerm{\Delta}{t}{a}$, then:
    \begin{equation}
      \label{eq:comp-term-pres}
      \denot{\TyIsoTerm{\Delta}{t}{a}}(\delta, \val{}) =
      \denot{\comp{\TyIsoTerm{\Delta}{t}{a}}}_{\gamma}%
      [(\compenv{\Delta,\delta}, \compsv{a}{\val{}})]
    \end{equation}
  \item If $~\TyIsoIso{\varnothing}{\omega}{a \leftrightarrow b}$, then:
    \begin{equation}
        \denot{%
          \TyIsoIso{\varnothing}{\omega}%
          {a \leftrightarrow b}%
          }(\val{1}, \val{2}) =
          \denot{%
          \comp{
          \TyIsoIso{\varnothing}{\omega}%
          {a \leftrightarrow b}
          }}_{\gamma}%
          [(\compsv{a}{\val{1}}, \compsv{b}{\val{2}})]
    \end{equation}
  \end{itemize}
\end{theorem}

This theorem can be proved by induction on the denotational semantics of ISO.
We show the proof for the $\kw{let}$ term in Appendix~\ref{section:proof}.
Other cases can be proved in a similar way.

\section{Optimizing Einsum for Sparsity and Expansion}\label{opmt}

Once we compile ISO programs to FGGs, we can use the existing FGG solver\footnote{FGGs: \url{https://github.com/diprism/fggs}} to simulate quantum computation.
However, large tensors encountered in probabilistic and quantum computation call for additional solver optimizations.
These optimizations preserve the semantics of tensor expressions, as we prove below.
They are also straightforward to implement (about 70 lines of Python added to the existing FGG solver).

\subsection{Saving Space and Time with $\expand$}\label{section:expand}

Consider a quantum circuit with $n$ qubits initialized to state $\ket{0}$, and the only operation in this circuit is a \rw{Hadamard} gate applied to the last qubit.
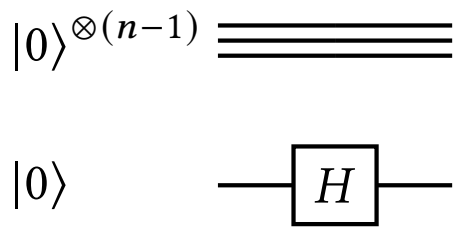
\[
  \begin{quantikz}
    &\ket{0}^{\otimes(n-1)~} & \qwbundle[alternate]{} & \qwbundle[alternate]{}\\
    &\ket{0}^{\hphantom{\otimes(n-1)~}} & \gate{H} & \qw
  \end{quantikz}
\]
Since this circuit is so simple, we expect that our solver could simulate quite a few qubits.
But it turns out that a naive solver---one that stores the state of $n$ qubits as $2^n$ numbers---cannot even simulate 30 qubits on a machine with 20~GB of RAM\@.
Intuitively, such circuits contain a lot of repetitive patterns.
We take advantage of those patterns to improve the simulator's performance.

Suppose we have a rank-$n$ tensor.
If every dimension has size~$2$, then the tensor contains $2^{n}$ elements, which is exponential in~$n$.
However, if we know that every element of the tensor is~$1$, then we do not need to allocate memory for $2^n$ elements.
We simply need to store the element~$1$ together with the tensor's shape.
Thus, using $\expand$ from FGGs, the tensor can be represented as $\expand(1[] \to x_{1}^{(2)}\cdots{x_{n}^{(2)}})$, which only takes $O(n)$ space.
Similarly, $\scatter$ can be used to store sparse tensors compactly.
In the representation $\expand(E[\vec{x}] \to \vec{z})$ or $\scatter(E[\vec{x}] \to e\dotsm)$ of a large tensor, we call the smaller tensor~$E$ the \emph{kernel}.

This representation saves not only memory but also computation.
When an $\expand$ed tensor is passed as an argument to $\einsum$, we can apply $\einsum$ to just the kernel of the tensor and expand the result to the actual shape.
For example, the tensor expression
\begin{equation}
  \einsum(\expand(4[] \to x^{(2)}y^{(3)})[x^{(2)}y^{(3)}],
  \expand(5[] \to y^{(3)})[y^{(3)}] \to
  x^{(2)}y^{(3)})
\end{equation}
computes the elementwise product between each row of the matrix $\left[
\begin{smallmatrix}
  4 & 4 & 4\\
  4 & 4 & 4
\end{smallmatrix}
\right]$ and the vector $[5\,5\,5]$, giving $\left[
\begin{smallmatrix}
  20 & 20 & 20\\
  20 & 20 & 20
\end{smallmatrix}
\right]
$.
Direct computation requires 6 multiplications.
If we apply $\einsum$ to the two kernels and then $\expand$, then since both kernels are scalars, we only need one multiplication:
\begin{equation}
  \expand(\einsum(4[], 5[] \to )[] \to x^{(2)}y^{(3)})
\end{equation}
In general, when an $\einsum$ is summation-free and some of its arguments are $\expand$ed tensors, we can swap the order of $\einsum$ and $\expand$.

On the other hand, if $\einsum$ is not summation-free, swapping $\einsum$ and $\expand$ is not enough.
For example, the tensor expression
\begin{equation}
  \einsum(\expand(4[] \to x^{(2)}y^{(3)})[x^{(2)}y^{(3)}],
  \expand(5[] \to y^{(3)})[y^{(3)}] \to
  x^{(2)})
\end{equation}
computes the dot product between each row of the matrix $\left[
\begin{smallmatrix}
  4 & 4 & 4\\
  4 & 4 & 4
\end{smallmatrix}
\right]$ and the vector $[5\,5\,5]$, giving $[60\,60]$.
If we do $\einsum$ first instead, by computing
\begin{equation}
  \expand(\einsum(4[], 5[] \to )[] \to x^{(2)}),
\end{equation}
we would get a different result $[20\,20]$.
We need to make up for it by multiplying a factor.

\subsection{Correctness of the Optimization}

Formally, our key optimization can be stated as follows.
We use $\einsum$ with two arguments here.
It can be generalized to $\einsum$ with more $\expand$ed arguments.

\begin{theorem}\label{thm:einsum-expand}
The operations $\einsum$ and $\expand$ commute up to a factor:
\begin{multline}
\llbracket
    \einsum\bigl(F[\vec{x}], \expand(E[\vec{y}] \to \vec{y}\vec{z})[\vec{y}\vec{z}] \to \vec{w}\bigr)
\rrbracket_{\gamma}
\\=
f\cdot
\llbracket
    \expand\bigl(\einsum(F[\vec{x}], E[\vec{y}] \to \vec{u})[\vec{u}] \to \vec{w}\bigr)
\rrbracket_{\gamma}\label{e:einsum-expand}
\end{multline}
where
\begin{align}
  \vec{u} &= \vec{w}\setminus(\vec{z}\setminus\vec{x}), &
  \vec{v} &= (\vec{z}\setminus\vec{x})\setminus\vec{w}, &
  f &= \textstyle\prod_{i \in \vec{v}}|i|
\end{align}
\end{theorem}

See Appendix~\ref{section:proof-einsum} for the proof of this theorem.

In \cref{fig:compiler}, we can see that $\einsum$ does not always take $\expand$ed tensors as arguments.
Sometimes, $\expand$ed tensors are nested in $\scatter$ed tensors (such as in the case for $\kw{inj}_{l}~t$).
Fortunately, just as our optimization reduces $\einsum$'s input from an $\expand$ed tensor to a smaller tensor, the FGG solver by \citet{chiangImplementation2025} already reduces $\einsum$'s input from a $\scatter$ed tensor to a smaller tensor. For instance, to store a diagonal matrix, it~represents the matrix as $\scatter(E[x] \to xx)$, where $E$ is the vector consisting of the elements on the diagonal.
To~multiply a matrix~$F$ by this diagonal matrix, the solver just multiplies $F$ by the vector~$E$:
\begin{equation}
    \llbracket \einsum(F[xy], \scatter(E[x] \to xx)[yz] \to xz) \rrbracket_{\gamma}
    = \llbracket \einsum(F[xz], E[z] \to xz) \rrbracket_{\gamma}.
\end{equation}
After this existing $\einsum$-$\scatter$ optimization takes effect, our new $\einsum$-$\expand$ optimization can then take over.

\subsection{Detecting Sparsity}

Whereas the $\einsum$-$\scatter$ and $\einsum$-$\expand$ optimizations benefit both probabilistic and quantum applications, other optimizations have a much bigger impact on quantum simulation than on probabilistic inference. The reversible nature of quantum computation calls for one specific optimization: detecting when multiplying two dense matrices yields an identity matrix. In particular, the pervasive Hadamard matrix $\frac{1}{\sqrt{2}}\bigl[\begin{smallmatrix}1&1\\1&-1\end{smallmatrix}\bigr]$ is its own inverse. Accordingly, after each $\einsum$ operation, our implementation checks for the specific dense result \smash[t]{$\bigl[\begin{smallmatrix}1&0\\0&1\end{smallmatrix}\bigr]$} and replaces it by the sparse equivalent $\scatter(\expand(1[]\to x)[x]\to xx)$.
This sparsity detection is performed after we get the result of an $\einsum$ at runtime.
Since this detection is restricted to a small matrix, it does not add much overhead to the solver.
When many intermediate results match this pattern, however, it can improve the simulation performance a lot, as shown by experiment~\labelcref{benchmark:dj-even} in \cref{fig:ablation}.

\section{Benchmark}\label{bench}

We perform benchmarks to answer three questions about our approach of quantum simulation by compiling through FGGs.
\begin{enumerate}
\item How does our approach compare in performance to other quantum simulators?
\item On what kinds of problems does our implementation perform well?
\item How do the components of our implementation (namely, the compiler and the solver optimizations) contribute to our performance?
\end{enumerate}
It turns out that our approach attains performance in the ballpark of several other quantum simulators.
It takes advantage of sparsity and preserves it through parallel composition.
Each component of our implementation improves performance on some benchmarks.
We first explain the design of our benchmarks in \cref{sec:benchmark-design}, then discuss the results in \cref{sec:benchmark-eval}.

\subsection{Benchmark Design}\label{sec:benchmark-design}

We measure the run times of the following simulators:
\begin{enumerate}
\item Compile ISO to FGGs directly as described in \cref{comp}, then use the FGG solver.\label{sim:fggi}

\item Compile ISO to PERPL\@, a~higher-level probabilistic programming language \cite{chiangExactRecursiveProbabilistic2023}, then use the existing compiler from PERPL to FGGs, followed by the FGG solver.
  We include this pipeline as another demonstration of reuse of existing probabilistic programming infrastructure.

\item Qiskit with the Aer state-vector simulator \cite{qiskit2024}.\footnote{We invoke \texttt{reverse\_qargs} to make sure the state vectors match other simulators.}

\item Cirq with the qsim state-vector simulator \cite{Cirq_Developers_2025}.

\item Quimb, a tensor network contraction simulator \cite{grayQuimbPythonPackage2018}.

\item qTorch, another tensor network contraction simulator \cite{friedQTorchQuantumTensor2018}.
Whereas the other simulators compute the entire state vector, qTorch only gives one measurement result at a time. Moreover, we observed no significant difference in performance across different measurement outcomes. Therefore, for qTorch simulating $n$ qubits, we run the experiment and multiply the run time by $2^{n}$, which is the number of possible measurement outcomes in the computational basis (or, the length of the state vector).
This choice is supported by the claim from the qTorch paper \cite{friedQTorchQuantumTensor2018} itself: ``To capture all the information of this final state of $n$ qubits, it generally requires $O(2^{n})$ repetitions of the algorithm.''
\end{enumerate}
We check that the results from all simulators match up to numerical precision.

We use each simulator to run the following experiments, which serve different purposes:
\begin{enumerate}[label=(\arabic*)]

\item Applying \rw{Hadamard} to the last qubit.\label{benchmark:had}
This is a simple circuit with a fixed number of operations.

\item The Deutsch-Jozsa algorithm for $n$ qubits, with the oracle function that checks if a natural number is even,\footnote{We also tried the zero oracle and obtained similar results.} where the oracle is simplified to a $C(X)$ on the last two qubits wrapped with two $X$ gates on the $(n-1)$th qubit \cite[Chapter~1]{nielsenQuantumComputationQuantum2010}.\label{benchmark:dj-even}
In this experiment, the number of operations grows linearly with~$n$.

\item The quantum part of Simon's algorithm on $n$ qubits, with $s = 2^{n} - 1$, where the oracle is represented as a dense $2^{n} \times 2^{n}$ unitary matrix \cite[Chapter~5]{nielsenQuantumComputationQuantum2010}.\label{benchmark:simon}
This experiment uses dense matrices as part of the algorithm.
For the two FGG-based simulators, we compile the oracle to an iso that enumerates all possible results.
Because qsim does not support dense matrices larger than $2^{6}\times2^{6}$,
we only run it up to 6 qubits.
And because qTorch accepts a limited version of QASM that only supports 1- and 2-qubit gates, we exclude it from this experiment.

\item $C^{n}(X)$ with $n$ control qubits decomposed to 1- and 2-qubit gates \cite[Chapter~4]{nielsenQuantumComputationQuantum2010}.\label{benchmark:mcx}
In~this experiment, the circuit size grows superlinearly with~$n$.
Because complex weights are required for larger~$n$, we cannot use PERPL to run this experiment, nor the next two.

\item Quantum Fourier Transform on $n$ qubits \cite[Chapter~5]{nielsenQuantumComputationQuantum2010}.\label{benchmark:qft}
This quantum algorithm has repeated circuit structures.

\item Grover's algorithm on $n$ qubits with $2^{n}$ entries and 0 as the required entry, where all gates are decomposed to 1- and 2-qubit gates \cite[Chapter~6]{nielsenQuantumComputationQuantum2010}.\label{benchmark:grover}
This quantum algorithm also has repeated circuit structures.

\item Bell state preparation of $n$ qubits \cite[Chapter~1]{nielsenQuantumComputationQuantum2010}.\label{benchmark:bell-state}

\item A parallel composition of \labelcref{benchmark:had} and \labelcref{benchmark:bell-state}: for $n + 10$ qubits, apply \rw{Hadamard} to qubit $n$ and prepare a Bell state on qubits $n+1$ through $n+10$.\label{benchmark:para-n-had-last-ten-qubit-bell}

\item Another parallel composition of \labelcref{benchmark:had} and \labelcref{benchmark:bell-state}: for $2n$ qubits, apply \rw{Hadamard} to qubit $n$ and prepare a Bell state on qubits $n+1$ through $2n$.\label{benchmark:para-n-had-last-n-bell}

\item A parallel composition of \labelcref{benchmark:had} and \labelcref{benchmark:dj-even}: for $2n$ qubits, apply \rw{Hadamard} to qubit $n$ and the Deutsch-Jozsa algorithm to qubits $n+1$ through $2n$.\label{benchmark:had-last-dj-even}
\end{enumerate}
We write the experiments in the same source language and compile them to ISO\@, Qiskit, Cirq, Quimb, and QASM programs while varying the number of qubits.

All experiments are performed on an Ubuntu 24.04 machine with a 2.50 GHz CPU of 48 cores and 503 GB of RAM\@.
All simulators are configured to use at most 24 threads (50\% of the core number) by their own settings.
For each qubit number, we limit the run time to 5 minutes and memory to 452.7 GB (90\% of the total memory).
We increase the number of qubits for a simulator until timeout or reaching the memory limit.
We configure the simulators as follows:
\begin{itemize}
\item FGG solver: based on PyTorch 2.10.0.
\item Qiskit: Qiskit 1.1.1, Aer 0.17.2, \texttt{StatevectorSimulator}, and transpile optimization level 2.
\item Cirq: Cirq 1.6.1, qsim 0.22.0; the maximum number of qubits allowed per fused gate is 2 (\texttt{max\_fused\_gate\_size=2}).
\item Quimb: Quimb 1.14.0, the \texttt{auto-hq} contraction path finding, the \texttt{to\_dense} contraction method.
\item qTorch: v1.1-3-gd564ae6, the \texttt{simple-stoch} contraction method.
\end{itemize}

\begin{figure}
\pgfplotsset{
  small,
  compat=1.5,
  every tick label/.append style={font=\tiny},
  label style={font=\footnotesize},
  title style={font=\footnotesize},
  ymode=log,
  table/col sep=comma,
  colormap/Paired-10,
  fggi/.style={
    color=Set1-A,
    mark=o,
    mark size=1.8pt,
    mark options={draw=Paired-F},
  },
  fggp/.style={
    color=Set1-B,
    mark=square,
    solid,
    mark size=1.8pt,
    mark options={draw=Paired-B,rotate=45},
  },
  qiskit/.style={
    color=Set1-C,
    mark=Mercedes star,
    solid,
    mark size=1.8pt,
  },
  cirq/.style={
    color=Set1-D,
    mark=x,
    solid,
    mark size=1.8pt,
  },
  qtorch/.style={
    color=Set1-E,
    mark=star,
    solid,
    mark size=1.8pt,
  },
  quimb/.style={
    color=Set1-H,
    mark=asterisk,
    solid,
    mark size=1.8pt,
  },
  gates/.style={
    color=gray,
    mark=none,
    dotted,
    semithick,
  },
}
\begin{center}
\begin{tikzpicture}
  \begin{groupplot}[group style={group size=2 by 3, group name=b1},
    group/horizontal sep=4em,
    group/vertical sep=5em]
    \nextgroupplot[
    xlabel={Number of qubits},
    ylabel={Running time (seconds)},
    axis y line*=left,
    title={\ref{benchmark:had} Hadamard to the last qubit},
    legend style={
      at={(0.5,-0.1)},
      anchor=north,legend columns=-1,
      font=\footnotesize,
      draw=none,
      legend columns=2,
      legend cell align={left},
    },
    legend to name=benchmark1,
    execute at end plot visualization/.append={%
      \xdef\savedxminhad{\pgfkeysvalueof{/pgfplots/xmin}}%
      \xdef\savedxmaxhad{\pgfkeysvalueof{/pgfplots/xmax}}%
      \xdef\savedyminhad{\pgfkeysvalueof{/pgfplots/ymin}}%
      \xdef\savedymaxhad{\pgfkeysvalueof{/pgfplots/ymax}}%
    },
    mark repeat=8,
    ]
    \addplot[fggi] table[x=qubit-number, y=had-last-qubit-allein_opt:iso-fgg-total] {benchmark/2026-10-11/had-last-qubit.csv};
    \addplot[fggp, mark phase=3] table[x=qubit-number, y=had-last-qubit-allein_opt:iso-perpl-fgg-total] {benchmark/2026-10-11/had-last-qubit.csv};
    \addplot[qiskit] table[x=qubit-number, y=had-last-qubit-avg:qiskit-simulation] {benchmark/2026-10-11/had-last-qubit.csv};
    \addplot[cirq] table[x=qubit-number, y=had-last-qubit-avg:qsim-simulation] {benchmark/2026-10-11/had-last-qubit.csv};
    \addplot[quimb] table[x=qubit-number, y=had-last-qubit-avg:quimb-simulation] {benchmark/2026-10-11/had-last-qubit.csv};

    \addlegendentry{FGG simulation with $\expand$ and $\bigl[\begin{smallmatrix}1&0\\0&1\end{smallmatrix}\bigr]$ opts};
    \addlegendentry{FGG simulation with $\expand$ and $\bigl[\begin{smallmatrix}1&0\\0&1\end{smallmatrix}\bigr]$ opts, through PERPL};
    \addlegendentry{Qiskit};
    \addlegendentry{Cirq};
    \addlegendentry{Quimb};
    \addlegendimage{qtorch};
    \addlegendentry{qTorch};
    \addlegendimage{gates};
    \addlegendentry{Number of gates};

    \nextgroupplot[
    xlabel={Number of qubits},
    axis y line*=left,
    title={\ref{benchmark:dj-even} Deutsch-Jozsa with the even oracle},
    mark repeat=4,
    execute at end plot visualization/.append={%
      \xdef\savedxmindje{\pgfkeysvalueof{/pgfplots/xmin}}%
      \xdef\savedxmaxdje{\pgfkeysvalueof{/pgfplots/xmax}}%
      \xdef\savedymindje{\pgfkeysvalueof{/pgfplots/ymin}}%
      \xdef\savedymaxdje{\pgfkeysvalueof{/pgfplots/ymax}}%
    },
    ]
    \addplot[fggi] table[x=qubit-number, y=deutsch-jozsa-is-even-simplified-allein_opt:iso-fgg-total] {benchmark/2026-10-11/deutsch-jozsa-is-even-simplified.csv};
    \addplot[fggp, mark phase=3] table[x=qubit-number, y=deutsch-jozsa-is-even-simplified-allein_opt:iso-perpl-fgg-total] {benchmark/2026-10-11/deutsch-jozsa-is-even-simplified.csv};
    \addplot[qiskit] table[x=qubit-number, y=deutsch-jozsa-is-even-simplified-avg:qiskit-simulation] {benchmark/2026-10-11/deutsch-jozsa-is-even-simplified.csv};
    \addplot[cirq] table[x=qubit-number, y=deutsch-jozsa-is-even-simplified-avg:qsim-simulation] {benchmark/2026-10-11/deutsch-jozsa-is-even-simplified.csv};
    \addplot[quimb] table[x=qubit-number, y=deutsch-jozsa-is-even-simplified-avg:quimb-simulation] {benchmark/2026-10-11/deutsch-jozsa-is-even-simplified.csv};

    \nextgroupplot[
    xlabel={Number of qubits},
    ylabel={Running time (seconds)},
    axis y line*=left,
    title={\ref{benchmark:simon} Simon's},
    mark repeat=1,
    execute at end plot visualization/.append={%
      \xdef\savedxminsimon{\pgfkeysvalueof{/pgfplots/xmin}}%
      \xdef\savedxmaxsimon{\pgfkeysvalueof{/pgfplots/xmax}}%
      \xdef\savedyminsimon{\pgfkeysvalueof{/pgfplots/ymin}}%
      \xdef\savedymaxsimon{\pgfkeysvalueof{/pgfplots/ymax}}%
    },
    ]
    \addplot[fggi] table[x=qubit-number, y=simon-allein_opt:iso-fgg-total] {benchmark/2026-10-11/simon.csv};
    \addplot[fggp] table[x=qubit-number, y=simon-allein_opt:iso-perpl-fgg-total] {benchmark/2026-10-11/simon.csv};
    \addplot[qiskit] table[x=qubit-number, y=simon-avg:qiskit-simulation] {benchmark/2026-10-11/simon.csv};
    \addplot[cirq] table[x=qubit-number, y=simon-avg:qsim-simulation] {benchmark/2026-10-11/simon.csv};
    \addplot[quimb] table[x=qubit-number, y=simon-avg:quimb-simulation] {benchmark/2026-10-11/simon.csv};

    \nextgroupplot[
    xlabel={Number of qubits},
    axis y line*=left,
    title={\ref{benchmark:mcx} Decomposed $C^{n}(X)$},
    mark repeat=2,
    execute at end plot visualization/.append={%
      \xdef\savedxminmcx{\pgfkeysvalueof{/pgfplots/xmin}}%
      \xdef\savedxmaxmcx{\pgfkeysvalueof{/pgfplots/xmax}}%
      \xdef\savedyminmcx{\pgfkeysvalueof{/pgfplots/ymin}}%
      \xdef\savedymaxmcx{\pgfkeysvalueof{/pgfplots/ymax}}%
    },
    ]
    \addplot[fggi] table[x=qubit-number, y=mcx-allein_opt:iso-fgg-total] {benchmark/2026-10-11/mcx.csv};
    \addplot[fggp] table[x=qubit-number, y=mcx-allein_opt:iso-perpl-fgg-total] {benchmark/2026-10-11/mcx.csv};
    \addplot[qiskit] table[x=qubit-number, y=mcx-avg:qiskit-simulation] {benchmark/2026-10-11/mcx.csv};
    \addplot[cirq] table[x=qubit-number, y=mcx-avg:qsim-simulation] {benchmark/2026-10-11/mcx.csv};
    \addplot[quimb] table[x=qubit-number, y=mcx-avg:quimb-simulation] {benchmark/2026-10-11/mcx.csv};

    \nextgroupplot[
    xlabel={Number of qubits},
    ylabel={Running time (seconds)},
    axis y line*=left,
    title={\ref{benchmark:qft} QFT},
    mark repeat=4,
    execute at end plot visualization/.append={%
      \xdef\savedxminqft{\pgfkeysvalueof{/pgfplots/xmin}}%
      \xdef\savedxmaxqft{\pgfkeysvalueof{/pgfplots/xmax}}%
      \xdef\savedyminqft{\pgfkeysvalueof{/pgfplots/ymin}}%
      \xdef\savedymaxqft{\pgfkeysvalueof{/pgfplots/ymax}}%
    },
    ]
    \addplot[fggi] table[x=qubit-number, y=qft-allein_opt:iso-fgg-total] {benchmark/2026-10-11/qft.csv};
    \addplot[fggp] table[x=qubit-number, y=qft-allein_opt:iso-perpl-fgg-total] {benchmark/2026-10-11/qft.csv};
    \addplot[qiskit] table[x=qubit-number, y=qft-avg:qiskit-simulation] {benchmark/2026-10-11/qft.csv};
    \addplot[cirq] table[x=qubit-number, y=qft-avg:qsim-simulation] {benchmark/2026-10-11/qft.csv};
    \addplot[quimb] table[x=qubit-number, y=qft-avg:quimb-simulation] {benchmark/2026-10-11/qft.csv};

    \nextgroupplot[
    xlabel={Number of qubits},
    axis y line*=left,
    title={\ref{benchmark:grover} Grover},
    mark repeat=2,
    execute at end plot visualization/.append={%
      \xdef\savedxmingrover{\pgfkeysvalueof{/pgfplots/xmin}}%
      \xdef\savedxmaxgrover{\pgfkeysvalueof{/pgfplots/xmax}}%
      \xdef\savedymingrover{\pgfkeysvalueof{/pgfplots/ymin}}%
      \xdef\savedymaxgrover{\pgfkeysvalueof{/pgfplots/ymax}}%
    },
    ]
    \addplot[fggi] table[x=qubit-number, y=grover-allein_opt:iso-fgg-total] {benchmark/2026-10-11/grover.csv};
    \addplot[fggp] table[x=qubit-number, y=grover-allein_opt:iso-perpl-fgg-total] {benchmark/2026-10-11/grover.csv};
    \addplot[qiskit] table[x=qubit-number, y=grover-avg:qiskit-simulation] {benchmark/2026-10-11/grover.csv};
    \addplot[cirq] table[x=qubit-number, y=grover-avg:qsim-simulation] {benchmark/2026-10-11/grover.csv};
    \addplot[quimb] table[x=qubit-number, y=grover-avg:quimb-simulation] {benchmark/2026-10-11/grover.csv};
  \end{groupplot}

  \begin{axis}[
    at={(b1 c1r1.south west)},
    anchor=south west,
    ymode=log,
    xmin=\savedxminhad,
    xmax=\savedxmaxhad,
    ymin=exp(\savedyminhad),
    ymax=exp(\savedymaxhad),
    mark repeat=4,
    hide axis,
    ]
    \addplot[qtorch] table[x=qubit-number, y=had-last-qubit-avg:qtorch-simulation-full] {benchmark/2026-10-11/had-last-qubit.csv};
  \end{axis}

  \begin{axis}[
    at={(b1 c2r1.south west)},
    anchor=south west,
    ymode=log,
    xmin=\savedxmindje,
    xmax=\savedxmaxdje,
    ymin=exp(\savedymindje),
    ymax=exp(\savedymaxdje),
    mark repeat=2,
    hide axis,
    ]
    \addplot[qtorch] table[x=qubit-number, y=deutsch-jozsa-is-even-simplified-avg:qtorch-simulation-full] {benchmark/2026-10-11/deutsch-jozsa-is-even-simplified.csv};
  \end{axis}

  \begin{axis}[
    at={(b1 c1r2.south west)},
    anchor=south west,
    ymode=log,
    xmin=\savedxminsimon,
    xmax=\savedxmaxsimon,
    ymin=exp(\savedyminsimon),
    ymax=exp(\savedymaxsimon),
    hide axis,
    ]
    \addplot[qtorch] table[x=qubit-number, y=simon-avg:qtorch-simulation-full] {benchmark/2026-10-11/simon.csv};
  \end{axis}

  \begin{axis}[
    at={(b1 c2r2.south west)},
    anchor=south west,
    ymode=log,
    xmin=\savedxminmcx,
    xmax=\savedxmaxmcx,
    ymin=exp(\savedyminmcx),
    ymax=exp(\savedymaxmcx),
    hide axis,
    ]
    \addplot[qtorch] table[x=qubit-number, y=mcx-avg:qtorch-simulation-full] {benchmark/2026-10-11/mcx.csv};
  \end{axis}

  \begin{axis}[
    at={(b1 c1r3.south west)},
    anchor=south west,
    ymode=log,
    xmin=\savedxminqft,
    xmax=\savedxmaxqft,
    ymin=exp(\savedyminqft),
    ymax=exp(\savedymaxqft),
    hide axis,
    ]
    \addplot[qtorch] table[x=qubit-number, y=qft-avg:qtorch-simulation-full] {benchmark/2026-10-11/qft.csv};
  \end{axis}

  \begin{axis}[
    at={(b1 c2r3.south west)},
    anchor=south west,
    ymode=log,
    xmin=\savedxmingrover,
    xmax=\savedxmaxgrover,
    ymin=exp(\savedymingrover),
    ymax=exp(\savedymaxgrover),
    hide axis,
    ]
    \addplot[qtorch] table[x=qubit-number, y=grover-avg:qtorch-simulation-full] {benchmark/2026-10-11/grover.csv};
  \end{axis}

  \begin{axis}[
    at={(b1 c1r1.south west)},
    anchor=south west,
    axis x line=none,
    axis y line*=right,
    ymode=log,
    ]
    \addplot[gates] table[x=qubit-number, y=gate-number] {benchmark/2026-10-11/had-last-qubit.csv};
  \end{axis}

  \begin{axis}[
    at={(b1 c2r1.south west)},
    anchor=south west,
    axis x line=none,
    axis y line*=right,
    ymode=log,
    ylabel={Number of gates},
    ]
    \addplot[gates] table[x=qubit-number, y=gate-number] {benchmark/2026-10-11/deutsch-jozsa-is-even-simplified.csv};
  \end{axis}

  \begin{axis}[
    at={(b1 c1r2.south west)},
    anchor=south west,
    axis x line=none,
    axis y line*=right,
    ymode=log,
    ]
    \addplot[gates] table[x=qubit-number, y=gate-number] {benchmark/2026-10-11/simon.csv};
  \end{axis}

  \begin{axis}[
    at={(b1 c2r2.south west)},
    anchor=south west,
    axis x line=none,
    axis y line*=right,
    ymode=log,
    ylabel={Number of gates},
    ]
    \addplot[gates] table[x=qubit-number, y=gate-number] {benchmark/2026-10-11/mcx.csv};
  \end{axis}

  \begin{axis}[
    at={(b1 c1r3.south west)},
    anchor=south west,
    axis x line=none,
    axis y line*=right,
    ymode=log,
    ]
    \addplot[gates] table[x=qubit-number, y=gate-number] {benchmark/2026-10-11/qft.csv};
  \end{axis}

  \begin{axis}[
    at={(b1 c2r3.south west)},
    anchor=south west,
    axis x line=none,
    axis y line*=right,
    ymode=log,
    ylabel={Number of gates},
    ]
    \addplot[gates] table[x=qubit-number, y=gate-number] {benchmark/2026-10-11/grover.csv};
  \end{axis}

  \path (b1 c1r3.south west) -- (b1 c2r3.south east) coordinate[midway] (lower_middle);
  \node[anchor=north, yshift=-2.5em] at (lower_middle) {\ref{benchmark1}};
\end{tikzpicture}
\end{center}
\vspace{-10pt}
\caption{Experiment results, comparing all simulators.
  For each plot, the $x$-axis shows the number of qubits, the left $y$-axis shows run time in seconds, and the right $y$-axis shows the number of gates.}
\label{fig:experiment}
\end{figure}
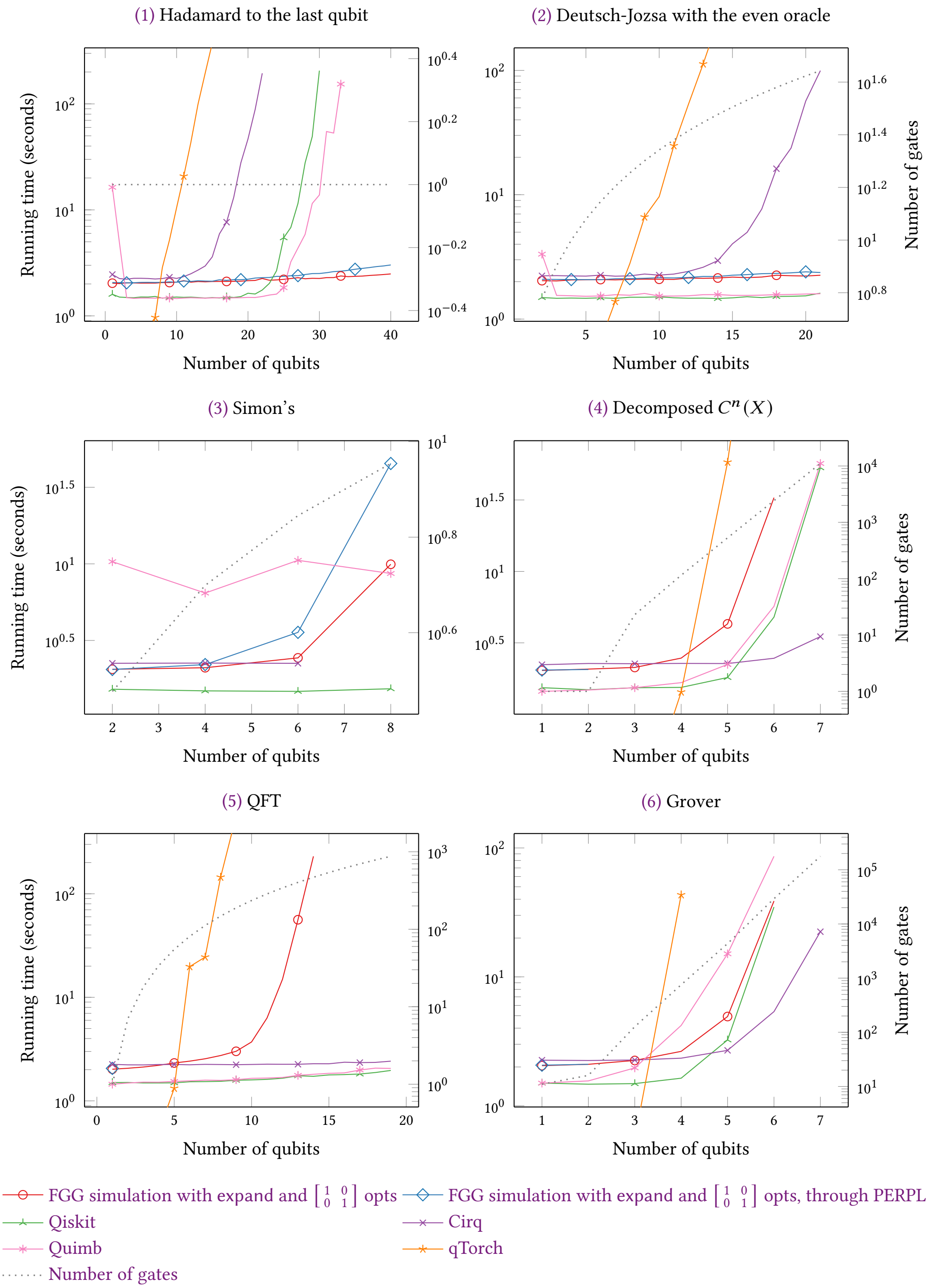

\begin{figure}
\pgfplotsset{
  small,
  compat=1.5,
  every tick label/.append style={font=\tiny},
  label style={font=\footnotesize},
  title style={font=\footnotesize},
  ymode=log,
  table/col sep=comma,
  colormap/Paired-10,
  fggi/.style={
    color=Set1-A,
    mark=o,
    mark size=1.8pt,
    mark options={draw=Paired-F},
  },
  fggp/.style={
    color=Set1-B,
    mark=square,
    solid,
    mark size=1.8pt,
    mark options={draw=Paired-B,rotate=45},
  },
  qiskit/.style={
    color=Set1-C,
    mark=Mercedes star,
    solid,
    mark size=1.8pt,
  },
  cirq/.style={
    color=Set1-D,
    mark=x,
    solid,
    mark size=1.8pt,
  },
  qtorch/.style={
    color=Set1-E,
    mark=star,
    solid,
    mark size=1.8pt,
  },
  quimb/.style={
    color=Set1-H,
    mark=asterisk,
    solid,
    mark size=1.8pt,
  },
  gates/.style={
    color=gray,
    mark=none,
    dotted,
    semithick,
  },
}
\begin{center}
\begin{tikzpicture}
  \begin{groupplot}[group style={group size=2 by 2, group name=b1},
    group/horizontal sep=4em,
    group/vertical sep=5em]
    \nextgroupplot[
    xlabel={Number of qubits},
    ylabel={Running time (seconds)},
    axis y line*=left,
    title={\ref{benchmark:bell-state} Bell state},
    legend style={
      at={(0.5,-0.1)},
      anchor=north,legend columns=-1,
      font=\footnotesize,
      draw=none,
      legend columns=2,
      legend cell align={left},
    },
    legend to name=benchmark-more,
    execute at end plot visualization/.append={%
      \xdef\savedxminhad{\pgfkeysvalueof{/pgfplots/xmin}}%
      \xdef\savedxmaxhad{\pgfkeysvalueof{/pgfplots/xmax}}%
      \xdef\savedyminhad{\pgfkeysvalueof{/pgfplots/ymin}}%
      \xdef\savedymaxhad{\pgfkeysvalueof{/pgfplots/ymax}}%
    },
    mark repeat=7,
    ]
    \addplot[fggi] table[x=qubit-number, y=bell-state-allein_opt:iso-fgg-total] {benchmark/2026-10-11/bell-state.csv};
    \addplot[fggp, mark phase=3] table[x=qubit-number, y=bell-state-allein_opt:iso-perpl-fgg-total] {benchmark/2026-10-11/bell-state.csv};
    \addplot[qiskit] table[x=qubit-number, y=bell-state-avg:qiskit-simulation] {benchmark/2026-10-11/bell-state.csv};
    \addplot[cirq] table[x=qubit-number, y=bell-state-avg:qsim-simulation] {benchmark/2026-10-11/bell-state.csv};
    \addplot[quimb] table[x=qubit-number, y=bell-state-avg:quimb-simulation] {benchmark/2026-10-11/bell-state.csv};

    \addlegendentry{FGG simulation with $\expand$ and $\bigl[\begin{smallmatrix}1&0\\0&1\end{smallmatrix}\bigr]$ opts};
    \addlegendentry{FGG simulation with $\expand$ and $\bigl[\begin{smallmatrix}1&0\\0&1\end{smallmatrix}\bigr]$ opts, through PERPL};
    \addlegendentry{Qiskit};
    \addlegendentry{Cirq};
    \addlegendentry{Quimb};
    \addlegendimage{qtorch};
    \addlegendentry{qTorch};
    \addlegendimage{gates};
    \addlegendentry{Number of gates};

    \nextgroupplot[
    xlabel={Number of qubits},
    axis y line*=left,
    title={\ref{benchmark:para-n-had-last-ten-qubit-bell} Hadamard in parallel with 10-qubit Bell},
    mark repeat=8,
    execute at end plot visualization/.append={%
      \xdef\savedxmindje{\pgfkeysvalueof{/pgfplots/xmin}}%
      \xdef\savedxmaxdje{\pgfkeysvalueof{/pgfplots/xmax}}%
      \xdef\savedymindje{\pgfkeysvalueof{/pgfplots/ymin}}%
      \xdef\savedymaxdje{\pgfkeysvalueof{/pgfplots/ymax}}%
    },
    ]
    \addplot[fggi] table[x=qubit-number, y=para-n-had-last-ten-qubit-bell-allein_opt:iso-fgg-total] {benchmark/2026-10-11/para-n-had-last-ten-qubit-bell.csv};
    \addplot[fggp] table[x=qubit-number, y=para-n-had-last-ten-qubit-bell-allein_opt:iso-perpl-fgg-total] {benchmark/2026-10-11/para-n-had-last-ten-qubit-bell.csv};
    \addplot[qiskit] table[x=qubit-number, y=para-n-had-last-ten-qubit-bell-avg:qiskit-simulation] {benchmark/2026-10-11/para-n-had-last-ten-qubit-bell.csv};
    \addplot[cirq] table[x=qubit-number, y=para-n-had-last-ten-qubit-bell-avg:qsim-simulation] {benchmark/2026-10-11/para-n-had-last-ten-qubit-bell.csv};
    \addplot[quimb] table[x=qubit-number, y=para-n-had-last-ten-qubit-bell-avg:quimb-simulation] {benchmark/2026-10-11/para-n-had-last-ten-qubit-bell.csv};

    \nextgroupplot[
    xlabel={Number of qubits},
    ylabel={Running time (seconds)},
    axis y line*=left,
    title={\ref{benchmark:para-n-had-last-n-bell} Hadamard in parallel with $n$-qubit Bell},
    mark repeat=4,
    execute at end plot visualization/.append={%
      \xdef\savedxminnhlnb{\pgfkeysvalueof{/pgfplots/xmin}}%
      \xdef\savedxmaxnhlnb{\pgfkeysvalueof{/pgfplots/xmax}}%
      \xdef\savedyminnhlnb{\pgfkeysvalueof{/pgfplots/ymin}}%
      \xdef\savedymaxnhlnb{\pgfkeysvalueof{/pgfplots/ymax}}%
    },
    ]
    \addplot[fggi] table[x=qubit-number, y=para-n-had-last-n-bell-allein_opt:iso-fgg-total] {benchmark/2026-10-11/para-n-had-last-n-bell.csv};
    \addplot[fggp] table[x=qubit-number, y=para-n-had-last-n-bell-allein_opt:iso-perpl-fgg-total] {benchmark/2026-10-11/para-n-had-last-n-bell.csv};
    \addplot[qiskit] table[x=qubit-number, y=para-n-had-last-n-bell-avg:qiskit-simulation] {benchmark/2026-10-11/para-n-had-last-n-bell.csv};
    \addplot[cirq] table[x=qubit-number, y=para-n-had-last-n-bell-avg:qsim-simulation] {benchmark/2026-10-11/para-n-had-last-n-bell.csv};
    \addplot[quimb] table[x=qubit-number, y=para-n-had-last-n-bell-avg:quimb-simulation] {benchmark/2026-10-11/para-n-had-last-n-bell.csv};

    \nextgroupplot[
    xlabel={Number of qubits},
    axis y line*=left,
    title={\ref{benchmark:had-last-dj-even} Hadamard in parallel with Deutsch-Jozsa},
    mark repeat=4,
    execute at end plot visualization/.append={%
      \xdef\savedxminhdj{\pgfkeysvalueof{/pgfplots/xmin}}%
      \xdef\savedxmaxhdj{\pgfkeysvalueof{/pgfplots/xmax}}%
      \xdef\savedyminhdj{\pgfkeysvalueof{/pgfplots/ymin}}%
      \xdef\savedymaxhdj{\pgfkeysvalueof{/pgfplots/ymax}}%
    },
    ]
    \addplot[fggi] table[x=qubit-number, y=had-last-dj-even-allein_opt:iso-fgg-total] {benchmark/2026-10-11/had-last-dj-even.csv};
    \addplot[fggp] table[x=qubit-number, y=had-last-dj-even-allein_opt:iso-perpl-fgg-total] {benchmark/2026-10-11/had-last-dj-even.csv};
    \addplot[qiskit] table[x=qubit-number, y=had-last-dj-even-avg:qiskit-simulation] {benchmark/2026-10-11/had-last-dj-even.csv};
    \addplot[cirq] table[x=qubit-number, y=had-last-dj-even-avg:qsim-simulation] {benchmark/2026-10-11/had-last-dj-even.csv};
    \addplot[quimb] table[x=qubit-number, y=had-last-dj-even-avg:quimb-simulation] {benchmark/2026-10-11/had-last-dj-even.csv};
  \end{groupplot}

  \begin{axis}[
    at={(b1 c1r1.south west)},
    anchor=south west,
    ymode=log,
    xmin=\savedxminhad,
    xmax=\savedxmaxhad,
    ymin=exp(\savedyminhad),
    ymax=exp(\savedymaxhad),
    mark repeat=4,
    hide axis,
    ]
    \addplot[qtorch] table[x=qubit-number, y=bell-state-avg:qtorch-simulation-full] {benchmark/2026-10-11/bell-state.csv};
  \end{axis}

  \begin{axis}[
    at={(b1 c2r1.south west)},
    anchor=south west,
    ymode=log,
    xmin=\savedxmindje,
    xmax=\savedxmaxdje,
    ymin=exp(\savedymindje),
    ymax=exp(\savedymaxdje),
    mark repeat=2,
    hide axis,
    ]
    \addplot[qtorch] table[x=qubit-number, y=para-n-had-last-ten-qubit-bell-avg:qtorch-simulation-full] {benchmark/2026-10-11/para-n-had-last-ten-qubit-bell.csv};
  \end{axis}

  \begin{axis}[
    at={(b1 c1r2.south west)},
    anchor=south west,
    ymode=log,
    xmin=\savedxminnhlnb,
    xmax=\savedxmaxnhlnb,
    ymin=exp(\savedyminnhlnb),
    ymax=exp(\savedymaxnhlnb),
    hide axis,
    ]
    \addplot[qtorch] table[x=qubit-number, y=para-n-had-last-n-bell-avg:qtorch-simulation-full] {benchmark/2026-10-11/para-n-had-last-n-bell.csv};
  \end{axis}

  \begin{axis}[
    at={(b1 c2r2.south west)},
    anchor=south west,
    ymode=log,
    xmin=\savedxminhdj,
    xmax=\savedxmaxhdj,
    ymin=exp(\savedyminhdj),
    ymax=exp(\savedymaxhdj),
    hide axis,
    ]
    \addplot[qtorch] table[x=qubit-number, y=had-last-dj-even-avg:qtorch-simulation-full] {benchmark/2026-10-11/had-last-dj-even.csv};
  \end{axis}

  \begin{axis}[
    at={(b1 c1r1.south west)},
    anchor=south west,
    axis x line=none,
    axis y line*=right,
    ymode=log,
    ]
    \addplot[gates] table[x=qubit-number, y=gate-number] {benchmark/2026-10-11/bell-state.csv};
  \end{axis}

  \begin{axis}[
    at={(b1 c2r1.south west)},
    anchor=south west,
    axis x line=none,
    axis y line*=right,
    ymode=log,
    ylabel={Number of gates},
    ]
    \addplot[gates] table[x=qubit-number, y=gate-number] {benchmark/2026-10-11/para-n-had-last-ten-qubit-bell.csv};
  \end{axis}

  \begin{axis}[
    at={(b1 c1r2.south west)},
    anchor=south west,
    axis x line=none,
    axis y line*=right,
    ymode=log,
    ]
    \addplot[gates] table[x=qubit-number, y=gate-number] {benchmark/2026-10-11/para-n-had-last-n-bell.csv};
  \end{axis}

  \begin{axis}[
    at={(b1 c2r2.south west)},
    anchor=south west,
    axis x line=none,
    axis y line*=right,
    ymode=log,
    ylabel={Number of gates},
    ]
    \addplot[gates] table[x=qubit-number, y=gate-number] {benchmark/2026-10-11/had-last-dj-even.csv};
  \end{axis}

  \path (b1 c1r2.south west) -- (b1 c2r2.south east) coordinate[midway] (lower_middle);
  \node[anchor=north, yshift=-2.5em] at (lower_middle) {\ref{benchmark-more}};
\end{tikzpicture}
\end{center}
\vspace{-10pt}
\caption{Experiment results, comparing all simulators (continued).
  For each plot, the $x$-axis shows the number of qubits, the left $y$-axis shows run time in seconds, and the right $y$-axis shows the number of gates.}
\label{fig:experiment-2}
\vspace{-10pt}
\end{figure}

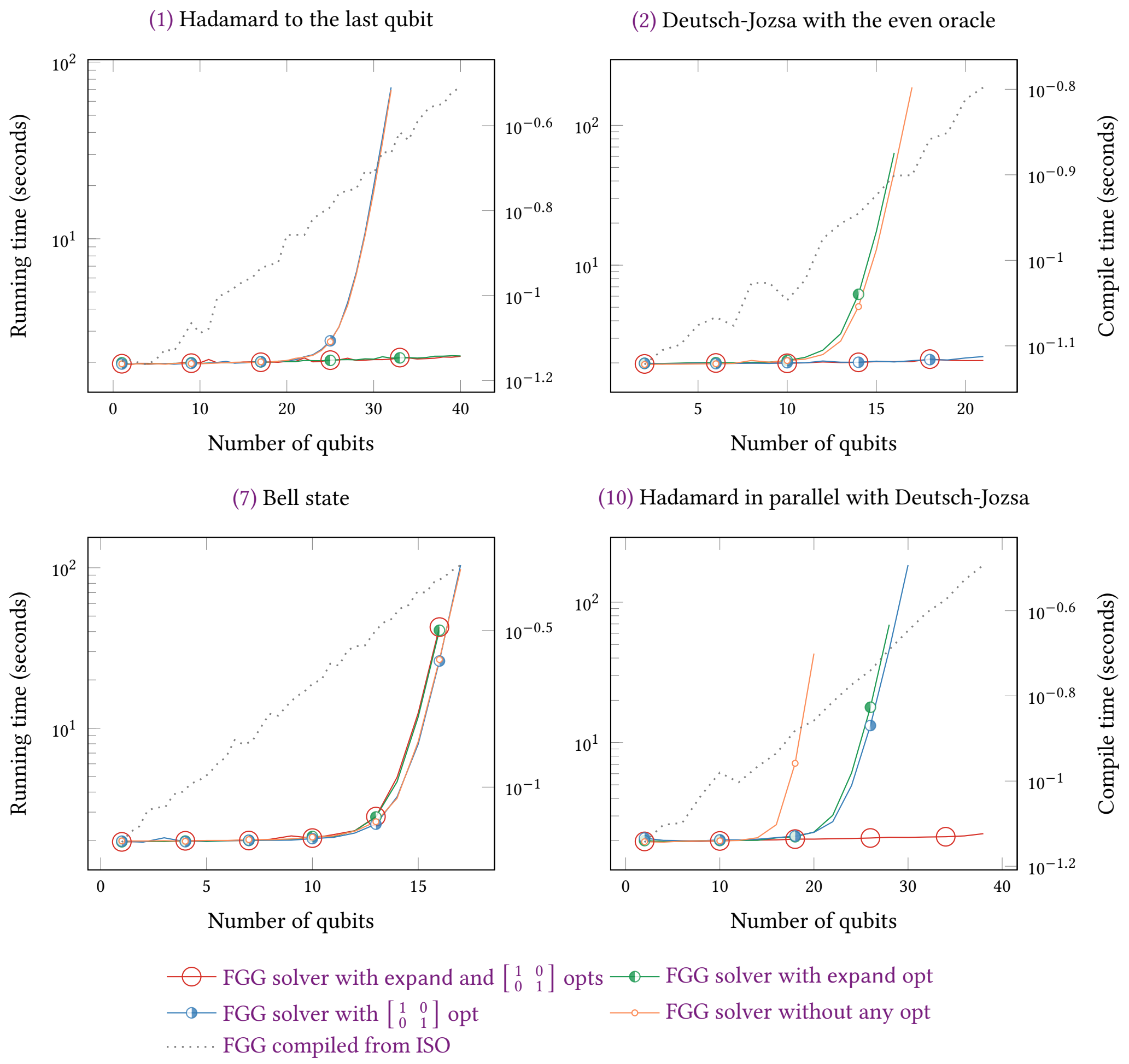
\begin{figure}
\pgfplotsset{
  small,
  compat=1.5,
  every tick label/.append style={font=\tiny},
  label style={font=\footnotesize},
  title style={font=\footnotesize},
  ytick pos=left,
  ymode=log,
  table/col sep=comma,
  clip marker paths=false,
  colormap/BuGn-9,
  cycle list/Set1-9,
  cycle multiindex* list={
    mark list*\nextlist
    Set1-9\nextlist
  },
  colormap/Paired-10,
  fggi-all/.style={
    color=RdYlGn-B,
    mark=o,
    mark size=3.2pt,
  },
  fggi-exp/.style={
    color=RdYlGn-N,
    mark=halfcircle*,
    solid,
    mark size=1.8pt,
    mark options={rotate=90, fill opacity=0.8, draw opacity=0.8},
  },
  fggi-unit/.style={
    color=Set1-B,
    mark=halfcircle*,
    solid,
    mark size=1.8pt,
    mark options={rotate=-90, fill opacity=0.8, draw opacity=0.8},
  },
  fggi-id/.style={
    color=RdYlGn-E,
    mark=halfcircle*,
    solid,
    mark size=1pt,
    mark options={fill=white,rotate=45},
  },
  fggi-comp/.style={
    color=gray,
    mark=none,
    dotted,
    semithick,
  },
  fggp-all/.style={
    color=Set1-B,
    mark=square,
    mark size=3.2pt,
    mark options={rotate=45},
  },
  fggp-exp/.style={
    color=PuOr-B,
    mark=halfsquare left*,
    solid,
    mark size=2.6pt,
  },
  fggp-unit/.style={
    color=PuOr-E,
    mark=halfsquare right*,
    solid,
    mark size=1.8pt,
  },
  fggp-id/.style={
    color=PuOr-N,
    mark=square,
    solid,
    mark size=1pt,
    mark options={fill=white,rotate=45},
  },
  fggp-comp/.style={
    color=YlGnBu-H,
    mark=x,
    solid,
    mark size=1.5pt,
  },
}
\begin{center}
\begin{tikzpicture}
  \begin{groupplot}[group style={group size=2 by 2, group name=b1},
    group/horizontal sep=4em,
    group/vertical sep=5em]
    \nextgroupplot[
    xlabel={Number of qubits},
    ylabel={Running time (seconds)},
    title={\ref{benchmark:had} Hadamard to the last qubit},
    legend style={
      at={(0.5,-0.1)},
      anchor=north,legend columns=-1,
      font=\footnotesize,
      draw=none,
      legend columns=2,
      legend cell align={left},
    },
    legend to name=benchmark2,
    mark repeat=8,
    ]
    \addplot[fggi-all] table[x=qubit-number, y=had-last-qubit-allein_opt:iso-fgg-simulation] {benchmark/2026-10-11/had-last-qubit.csv};
    \addplot[fggi-exp] table[x=qubit-number, y=had-last-qubit-ein_opt:iso-simulation] {benchmark/2026-10-11/had-last-qubit.csv};
    \addplot[fggi-unit] table[x=qubit-number, y=had-last-qubit-id_detect:iso-fgg-simulation] {benchmark/2026-10-11/had-last-qubit.csv};
    \addplot[fggi-id] table[x=qubit-number, y=had-last-qubit-no_opt:iso-fgg-simulation] {benchmark/2026-10-11/had-last-qubit.csv};
    \addlegendentry{FGG solver with $\expand$ and $\bigl[\begin{smallmatrix}1&0\\0&1\end{smallmatrix}\bigr]$ opts};
    \addlegendentry{FGG solver with $\expand$ opt};
    \addlegendentry{FGG solver with $\bigl[\begin{smallmatrix}1&0\\0&1\end{smallmatrix}\bigr]$ opt};
    \addlegendentry{FGG solver without any opt};
    \addlegendimage{fggi-comp};
    \addlegendentry{FGG compiled from ISO};

    \nextgroupplot[
    xlabel={Number of qubits},
    title={\ref{benchmark:dj-even} Deutsch-Jozsa with the even oracle},
    mark repeat=4,
    ]
    \addplot[fggi-all] table[x=qubit-number, y=deutsch-jozsa-is-even-simplified-allein_opt:iso-fgg-simulation] {benchmark/2026-10-11/deutsch-jozsa-is-even-simplified.csv};
    \addplot[fggi-exp] table[x=qubit-number, y=deutsch-jozsa-is-even-simplified-ein_opt:iso-fgg-simulation] {benchmark/2026-10-11/deutsch-jozsa-is-even-simplified.csv};
    \addplot[fggi-unit] table[x=qubit-number, y=deutsch-jozsa-is-even-simplified-id_detect:iso-fgg-simulation] {benchmark/2026-10-11/deutsch-jozsa-is-even-simplified.csv};
    \addplot[fggi-id] table[x=qubit-number, y=deutsch-jozsa-is-even-simplified-no_opt:iso-fgg-simulation] {benchmark/2026-10-11/deutsch-jozsa-is-even-simplified.csv};

    \nextgroupplot[
    xlabel={Number of qubits},
    ylabel={Running time (seconds)},
    title={\ref{benchmark:bell-state} Bell state},
    mark repeat=3,
    ]
    \addplot[fggi-all]  table[x=qubit-number, y=bell-state-allein_opt:iso-fgg-simulation] {benchmark/2026-10-11/bell-state.csv};
    \addplot[fggi-exp]  table[x=qubit-number, y=bell-state-ein_opt:iso-fgg-simulation] {benchmark/2026-10-11/bell-state.csv};
    \addplot[fggi-unit]  table[x=qubit-number, y=bell-state-id_detect:iso-fgg-simulation] {benchmark/2026-10-11/bell-state.csv};
    \addplot[fggi-id]   table[x=qubit-number, y=bell-state-no_opt:iso-fgg-simulation] {benchmark/2026-10-11/bell-state.csv};

    \nextgroupplot[
    xlabel={Number of qubits},
    title={\ref{benchmark:had-last-dj-even} Hadamard in parallel with Deutsch-Jozsa},
    mark repeat=4,
    ]
    \addplot[fggi-all]  table[x=qubit-number, y=had-last-dj-even-allein_opt:iso-fgg-simulation] {benchmark/2026-10-11/had-last-dj-even.csv};
    \addplot[fggi-exp]  table[x=qubit-number, y=had-last-dj-even-ein_opt:iso-fgg-simulation] {benchmark/2026-10-11/had-last-dj-even.csv};
    \addplot[fggi-unit]  table[x=qubit-number, y=had-last-dj-even-id_detect:iso-fgg-simulation] {benchmark/2026-10-11/had-last-dj-even.csv};
    \addplot[fggi-id]   table[x=qubit-number, y=had-last-dj-even-no_opt:iso-fgg-simulation] {benchmark/2026-10-11/had-last-dj-even.csv};
  \end{groupplot}

  \begin{axis}[
    at={(b1 c1r1.south west)},
    anchor=south west,
    axis x line=none,
    axis y line*=right,
    ymode=log,
    ]
    \addplot[fggi-comp] table[x=qubit-number, y=had-last-qubit-avg:iso-to-fgg] {benchmark/2026-10-11/had-last-qubit.csv};
  \end{axis}

  \begin{axis}[
    at={(b1 c2r1.south west)},
    anchor=south west,
    axis x line=none,
    axis y line*=right,
    ymode=log,
    ylabel={Compile time (seconds)},
    ]
    \addplot[fggi-comp] table[x=qubit-number, y=deutsch-jozsa-is-even-simplified-avg:iso-to-fgg] {benchmark/2026-10-11/deutsch-jozsa-is-even-simplified.csv};
  \end{axis}

  \begin{axis}[
    at={(b1 c1r2.south west)},
    anchor=south west,
    axis x line=none,
    axis y line*=right,
    ymode=log,
    ]
    \addplot[fggi-comp] table[x=qubit-number, y=bell-state-avg:iso-to-fgg] {benchmark/2026-10-11/bell-state.csv};
  \end{axis}

  \begin{axis}[
    at={(b1 c2r2.south west)},
    anchor=south west,
    axis x line=none,
    axis y line*=right,
    ymode=log,
    ylabel={Compile time (seconds)},
    ]
    \addplot[fggi-comp] table[x=qubit-number, y=had-last-dj-even-avg:iso-to-fgg] {benchmark/2026-10-11/had-last-dj-even.csv};
  \end{axis}
  \path (b1 c1r2.south west) -- (b1 c2r2.south east) coordinate[midway] (lower_middle);
  \node[anchor=north, yshift=-2.5em] at (lower_middle) {\ref{benchmark2}};
\end{tikzpicture}
\end{center}
\vspace{-10pt}
\caption{Simulation and compilation times of our FGG-based simulators, comparing different sets of optimizations.
  For each plot, the $x$-axis shows the number of qubits, the left $y$-axis shows run time in seconds, and the right $y$-axis shows compilation time in seconds.}
\label{fig:ablation}
\vspace{-10pt}
\end{figure}

\subsection{Benchmark Evaluation}\label{sec:benchmark-eval}

\subsubsection{Comparison among simulators}
The run times of all simulators are shown in \cref{fig:experiment,fig:experiment-2}.
Each plot shows one experiment and compares different simulators.
For each plot, the $x$-axis is the number of qubits, the left $y$-axis is the run time in seconds (log scale), and the right $y$-axis (used only by the dotted line) is the total number of gates (log scale).
The run times shown for our two FGG-based simulators include both compiling from ISO and solving the resulting FGGs.

There is no single simulator that achieves the best performance across all experiments and sizes.
Overall, our FGG-based simulators attain performance in the ballpark of the other simulators.
For the single Hadamard gate (experiment \labelcref{benchmark:had}), both FGG implementations can simulate up to 40 qubits.
For the Deutsch-Jozsa algorithm (experiment~\labelcref{benchmark:dj-even}), the FGG simulator is faster than qsim but slower than Qiskit and Quimb.
For Simon's algorithm (experiment~\labelcref{benchmark:simon}), Qiskit, Quimb, and the FGG implementations can simulate up to 8 qubits.
For $C^{n}(X)$ and Grover (experiments~\labelcref{benchmark:mcx,benchmark:grover}), qsim achieves the best performance when the program size increases.
For QFT and Bell state preparation (experiments~\labelcref{benchmark:qft,benchmark:bell-state}), Qiskit, qsim, and Quimb outperform FGG when the program size increases.
For QFT, we also noticed that Qiskit introduces a global phase shift when the number of qubits exceeds 3.

Experiments \labelcref{benchmark:para-n-had-last-ten-qubit-bell,benchmark:para-n-had-last-n-bell,benchmark:had-last-dj-even} are parallel compositions of experiment \labelcref{benchmark:had}, where our FGG-based simulators performed the best at high qubit counts, with experiments \labelcref{benchmark:dj-even,benchmark:bell-state}, where they did not.
Parallel composition improved the standing of our FGG-based simulators relative to other simulators, so~much so that our simulators outperform others at high qubit counts in experiments~\labelcref{benchmark:para-n-had-last-ten-qubit-bell,benchmark:had-last-dj-even}.
The improvement is due to the fact that the FGG implementations can take advantage of the programs' patterned sparsity and preserve it when the programs are composed in parallel.

\subsubsection{Ablation of optimizations}
To further study which components of our implementation contribute to the performance, we run the experiments with different optimizations and record the run times of the FGG solver and our compiler separately.
We compile the same ISO source programs to FGGs directly.
Then for each compilation result, we run the FGG solver with
\begin{enumerate}
\item both the $\expand$ equation rewriting and sparsity detection enabled,
\item the $\expand$ equation rewriting enabled and sparsity detection disabled,
\item the $\expand$ equation rewriting disabled and sparsity detection enabled, and
\item no optimization enabled (but optimizations in the original FGG solver remain active).
\end{enumerate}
This gives us four FGG solver run times and one compilation time.
Four representative results are shown in \cref{fig:ablation}.
The plots correspond to experiments~\labelcref{benchmark:had,benchmark:dj-even,benchmark:bell-state,benchmark:had-last-dj-even} as shown in \cref{fig:experiment,fig:experiment-2}.
The ablation experiments for the other benchmarks show similar patterns to these four plots.
The compilation path through PERPL also shows similar results.
For each plot, the $x$-axis is the number of qubits, the left $y$-axis is the run time in seconds (log scale), and the right $y$-axis (used only by the dotted line) is the compilation time (log scale).

These experiments show that our optimizations are good at preserving sparsity under parallel composition.
Experiments~\labelcref{benchmark:had,benchmark:dj-even} show that both $\expand$ equation rewriting and sparsity detection can improve performance when the programs have the corresponding structure.
The structure $\expand(\dots\einsum\dots)$ generated by $\data{comp}{}{}$ helps the FGG solver rewrite $\expand$ expressions to smaller sizes.
The inlining of single-variable \kw{let} expressions in $\data{comp}{}{}$ helps the FGG solver detect potentially sparse structures.
For example, by substitution, $\kw{let}~y = \rw{Hadamard}~x~\kw{in}~\kw{let}~z = \rw{Hadamard}~y~\kw{in}~z$ is first converted to $\kw{let}~z = \rw{Hadamard}~(\rw{Hadamard}~x)~\kw{in}~z$, then it is compiled to an FGG tensor.
Since two consecutive $\rw{Hadamard}$s are equivalent to identity, the sparsity detection can reduce it to a sparse tensor.
Once such structures are recognized, tensor products of these sparse tensors can remain sparse.
In the best case, a tensor of size $2^{n}\times 2^{n}$ can be represented by only 1 numerical element in memory, thus improving the overall memory usage and run times exponentially.
Experiment~\labelcref{benchmark:had-last-dj-even} shows that the two optimizations together can make some programs outperform any single optimization.
When programs don't have the structure that the FGG solver recognizes, however, even if the final result is sparse, the solver might still need exponential time when the program size increases.
In this case, neither optimization helps the performance, as shown in experiment~\labelcref{benchmark:bell-state}.

Compilation becomes a significant bottleneck in experiments~\labelcref{benchmark:simon,benchmark:mcx,benchmark:qft,benchmark:grover}, and a sole bottleneck in experiment~\labelcref{benchmark:simon}.
This is unsurprising because encoding a dense $2^n \times 2^n$ unitary matrix is inherently exponential in~$n$.
Our compilers need to traverse programs and generate a JSON file representing the FGGs, during which process they may print parts of a large program multiple times.
In experiments~\labelcref{benchmark:mcx,benchmark:qft,benchmark:grover}, the significant compile time imposes one bottleneck, but another bottleneck is that the FGG solver runs out of memory.


\section{Related Work}\label{related}

The design of quantum programming languages is an active research area \cite{selingerQuantumProgrammingLanguage2004,yingDefiningQuantumControl2012,greenQuipperScalableQuantum2013,paykinQWIRECoreLanguage2017,bichselSilqHighlevelQuantum2020,voichickQunityUnifiedLanguage2023,daveCombiningQuantumClassical2025}.
As a representative, ISO models pure quantum programs with a linear type system and a syntax that follows pattern matching in functional programming.
Other languages explore alternative design choices that can make quantum programming more accessible.
It remains to be seen what is the best way to do high level quantum programming, design quantum control flow, and combine quantum programming with classical programming.

Various probabilistic graphical models have been proposed \cite{kollerProbabilisticGraphicalModels2009}.
As described in \cref{tgt}, FGGs are picked as a representative probabilistic graphical model.
Other graphical models can also support our experiments if they satisfy our criteria.

Below we give a brief introduction to the interaction between probabilistic programming and quantum programming.

\subsubsection*{Generalizing classical probability models to quantum ones}
\citet{tucciQuantumBayesianNets1995} extends Bayesian networks to Quantum Bayesian networks, where probability amplitudes can be negative or complex.
\citet{yeangProbabilisticGraphicalModel2010} defines a probabilistic graphical model for quantum systems.
In this model, vertices are quantum states and edges are possible operations like unitary gates or measurements.
\citet{warmuthBayesianProbabilityCalculus2014} develop a probability calculus that generalizes the classical one to describe density matrices.
\citet{inoueQuantumProgrammingQuantum2024} takes the idea of negative probability from \citet{feynmanSimulatingPhysicsComputers1982} and designs a universal quantum programming language with a two-layer probability model to support both classical probabilities and quantum amplitudes, which may be negative.
These models study theoretical quantum generalizations of probabilistic models, but they address little about implementation, verification, or reuse of existing probabilistic infrastructures.

\subsubsection*{Extending quantum programming languages with probabilistic operations}
\citet{yuanProbabilisticInferenceQuantum2021} extends the quantum \rw{while}-language designed by \citet{yingFloydHoareLogic2012} with a probabilistic operator \rw{observe} and supports Bayesian inference for quantum programs.
But they only give a prototype and don't benchmark the performance or verify the language's semantics.
\citet{gehnenBayesianInferenceQuantum2025} make a similar extension and discuss the semantics of such a language through weakest precondition.
But they don't give an implementation.

\subsubsection*{Probabilistic simulation of quantum computation}
\citet{obeidEvaluatingProbabilisticProgramming2019} simulate an EPR experiment using different probabilistic programming languages.
This experiment demonstrates reuse of probabilistic programming infrastructure for EPR simulation, but does not address larger programs, performance, or formal semantics.
\citet{huangLogicalAbstractionsNoisy2021} compile noisy quantum circuits into complex-valued Bayesian networks and use conjunctive normal form for optimization.
They benchmark the performance of their simulator but don't give a formal semantics or prove the correctness of the compilation.
Machine learning models, such as Transformer \cite{PhysRevA.104.032610,langeTransformerNeuralNetworks2025}, are also used to encode quantum states and simulate quantum circuits, though their concern is less about the semantics and correctness.

\section{Future Work}\label{future}


Our FGG solver supports auto-differentiation.
It remains to be seen how to apply this feature to more quantum programs.

Our FGG solver already uses the PyTorch CPU parallelization.
It remains to be seen if using PyTorch GPU parallelization can improve the performance.

PyTorch limits tensor ranks to a small number.
For example, the operation \texttt{torch.sum} supports tensors up to rank 64 as of our test.\footnote{\url{https://github.com/pytorch/pytorch/issues/23159}}
If ISO's typing context contains more variables than this limit, PyTorch simply rejects the compiled FGGs.
Thus, the number of qubits that can be simulated is constrained by this limit.
If we bypass this limit, we might be able to simulate more qubits.

In addition to symmetric pattern matching, ISO also supports higher-order isos and structural recursion on recursive data types like natural numbers and lists \cite{sabrySymmetricPatternmatchingQuantum2018}.
Probabilistic programming languages like PERPL \cite{chiangExactRecursiveProbabilistic2023} also support higher order functions, recursive data types, and recursive functions.
PERPL uses defunctionalization and refunctionalization to eliminate recursive data types.
Similar techniques should be applicable to higher-order and recursive isos.

\section{Conclusion}\label{conc}

We compile a high-level quantum language ISO to a probabilistic computation representation FGGs and simulate quantum programs using the existing FGG solver.
We enhance the FGG solver to rewrite tensor expressions so as to perform $\einsum$ on smaller tensors.
We prove that the compilation and the rewriting both preserve denotational semantics.
We compare our implementations with other simulators on various quantum programs.
The experiment results show that our approach achieves performance comparable to other quantum simulators, and the rewriting improves our simulation performance for quantum programs with repeated patterns.
Our work establishes a connection between quantum programming and probabilistic programming, demonstrating benefits of reusing probabilistic programming techniques for quantum simulation.

\section{Data-Availability Statement}

We plan to submit an artifact.
The artifact will include the following parts:
\begin{enumerate}
\item The source code of our implementations, including the compilers and the FGG solver.
\item The scripts to generate quantum programs and run the experiments.
\item The performance data collected from running our experiments.
\item Detailed instructions on how to reproduce our results.
\end{enumerate}

\bibliographystyle{ACM-Reference-Format}
\bibliography{main}

\clearpage
\appendix

\section{Denotational Semantics of the Remaining ISO Constructs}\label{section:denot-iso-more}

The denotational semantics of ISO patterns, pure values, and extended values are shown in \cref{fig:iso-denote-more}.
We use $\denotpat{\_}$, $\denotpv{\_}$, and $\denotext{\_}$ for patterns, pure values, and extended values respectively.
They follow the definitions in \crefrange{denot:iso-t-unit}{denot:iso-t-scale}, all denoting a function that takes an environment and a semantic value and returns a weight.
For $\kw{let}~p_{1}~\kw{=}~\omega~{p_{2}}~\kw{in}~e$, we iterate through the semantic values that $p_{1}$ and $p_{2}$ can take, and the environments corresponding to the typing context of $p_{2}$.
\cref{denot:iso-ext-let} can be thought of as a combination of \cref{denot:iso-t-app,denot:iso-t-let}.

\begin{figure}[htp]
\raggedright{Patterns}
\begin{alignat}{2}
  &\denotpat{\TyIsoPattern{\varnothing}{\kw{()}}{\mathbbold{1}}}(\varnothing, \kw{()})
  &&=
  1\\
  &\denotpat{\TyIsoPattern{x:a}{x}{a}}(\{(x,\val{}')\}, \val{})
  &&=
  \begin{cases}
    1 & \text{if}~\val{}' = \val{}\\
    0 & \text{otherwise}
  \end{cases}\\
  &\denotpat{\TyIsoPattern{\Delta_{1};\Delta_{2}}%
    {\langle{p_{1}, p_{2}}\rangle}%
    {a \otimes b}}(\delta_{1} \cup \delta_{2}, \langle{\val{1}, \val{2}}\rangle)
  &&=
  \denotpat{\TyIsoPattern{\Delta_{1}}%
    {p_{1}}%
    {a}}(\delta_{1}, \val{1}) \cdot
    \denotpat{\TyIsoPattern{\Delta_{2}}%
    {p_{2}}%
    {b}}(\delta_{2}, \val{2})
\end{alignat}
\raggedright{Pure Values}
\begin{alignat}{2}
  &\denotpv{\TyIsoVal{\varnothing}{\kw{()}}{\mathbbold{1}}}(\varnothing, \kw{()})
  &&=
  1\\
  &\denotpv{\TyIsoVal{x:a}{x}{a}}(\{(x,\val{}')\}, \val{})
  &&=
  \begin{cases}
    1 & \text{if}~\val{}' = \val{}\\
    0 & \text{otherwise}
  \end{cases}\\
  &\denotpv{\TyIsoVal{\Delta}{\kw{inj}_{l}~t}{a \oplus b}}(\delta, \kw{inj}_{i}~\val{})
  &&=
  \begin{cases}
    \denotpv{\TyIsoVal{\Delta}{t}{a}}(\delta, \val{}) & \text{if}~i=l \\
    0 & \text{if}~i=r \\
  \end{cases}\\
  &\denotpv{\TyIsoVal{\Delta}{\kw{inj}_{r}~t}{a \oplus b}}(\delta, \kw{inj}_{i}~\val{})
  &&=~
  \begin{cases}
    \denotpv{\TyIsoVal{\Delta}{t}{b}}(\delta, \val{}) & \text{if}~i=r \\
    0 & \text{if}~i=l \\
  \end{cases}\\
  &\denotpv{\TyIsoVal{\Delta_{1};\Delta_{2}}%
    {\langle{v_{1}, v_{2}}\rangle}%
    {a \otimes b}}(\delta_{1} \cup \delta_{2}, \langle{\val{1}, \val{2}}\rangle)
  &&=
  \denotpv{\TyIsoVal{\Delta_{1}}%
    {v_{1}}%
    {a}}(\delta_{1}, \val{1}) \cdot
    \denotpv{\TyIsoVal{\Delta_{2}}%
    {v_{2}}%
    {b}}(\delta_{2}, \val{2})
\end{alignat}
\raggedright{Extended Values}
\begin{alignat}{2}
  &\denotext{\TyIsoExt{\Delta}{v}{a}}(\delta, \val{})
  &&=
  \denotpv{\TyIsoVal{\Delta}{v}{a}}(\delta, \val{})\\
  &\denotext{\TyIsoExt{\Delta}{e_{1} + e_{2}}{a}}(\delta, \val{})
  &&=
  \denotext{\TyIsoExt{\Delta}{e_{1}}{a}}(\delta, \val{}) +
    \denotext{\TyIsoExt{\Delta}{e_{2}}{a}}(\delta, \val{})\\
  &\denotext{\TyIsoExt{\Delta}{\alpha{e}}{a}}(\delta, \val{})
  &&=
  \alpha \cdot \denotext{\TyIsoExt{\Delta}{e}{a}}(\delta, \val{})\\
  \label{denot:iso-ext-let}
  &\denotext{\TyIsoExt{\Delta_{1};\Delta_{2}}{\kw{let}~p_{1}~\kw{=}~\omega~{p_{2}}~\kw{in}~e}{b}}%
    (\delta_{1} \cup \delta_{2}, \val{}')\notag\\
  &\qquad=
  \sum_{\substack{\delta_{2}' \in \denot{\Delta_{2}'}\\\val{a} \in \denot{a}\\\val{a'} \in \denot{a'}}}{
  \mathrlap{%
  \begin{aligned}
    &\denot{%
      \TyIsoIso{\varnothing}%
      {\omega}%
      {a \leftrightarrow a'}}%
      (\val{a}, \val{a'}) \cdot
      \denotpat{\TyIsoPattern{\Delta_{1}}{p_{2}}{a}}(\delta_{1}, \val{a})
      \cdot\\
    &\denotpat{\TyIsoPattern{\Delta_{2}'}{p_{1}}{a}}%
      (\delta_{2}', \val{a'}) \cdot\\
    &\denotext{\TyIsoTerm{\Delta_{2};\Delta_{2}'}{e}{b}}%
      (\delta_{2} \cup \delta_{2}', \val{}')
  \end{aligned}
  }}
\end{alignat}
\caption{The denotation of ISO patterns, pure values, and extended values}
\label{fig:iso-denote-more}
\end{figure}

\section{Compilation of the Remaining ISO Constructs}\label{section:comp-iso-more}

\begin{figure}
\centering
\begin{minipage}[t]{0.48\textwidth}
\begin{align*}
    &\data{comp}{\mathrm{p}}{} ::~ \data{Pattern}{iso}{} \to \data{Tensor}{}{} \\
    &\comppat{p} ={} \comp{\comppt{p}}\\
    \eqline
    &\data{comp}{\mathrm{v}}{} ::~ \data{Value}{iso}{} \to \data{Tensor}{}{} \\
    &\compv{v} ={} \comp{\compvt{v}}\\
    \eqline
    &\data{comp}{\mathrm{e}}{} ::~ \data{ExtValue}{iso}{} \to \data{Tensor}{}{}\\
    &\compext{e} = \comp{\compet{e}}\\
    \eqline\\
    &\data{comp}{\mathrm{p}}{\mathrm{t}} ::~ \data{Pattern}{iso}{} \to \data{Term}{iso}{}\\
    &\comppt{\kw{()}} ={} \kw{()}\\
    &\comppt{x} ={} x\\
    &\comppt{\langle{p_{1}, p_{2}}\rangle} ={} \langle{\comppt{p_{1}}, \comppt{p_{2}}}\rangle\\
\end{align*}
\end{minipage}\hfill
\begin{minipage}[t]{0.48\textwidth}
\begin{align*}
    &\data{comp}{\mathrm{v}}{\mathrm{t}} :: \data{Value}{iso}{} \to \data{Term}{iso}{}\\
    &\compvt{\kw{()}} = \kw{()}\\
    &\compvt{x} = x\\
    &\compvt{\kw{inj}_{l}~v} = \kw{inj}_{l}~\compvt{v}\\
    &\compvt{\kw{inj}_{r}~v} = \kw{inj}_{r}~\compvt{v}\\
    &\compvt{\langle{v_{1}, v_{2}}\rangle} = \langle{\compvt{v_{1}}, \compvt{v_{2}}}\rangle\\
    \eqline
    &\data{comp}{\mathrm{e}}{\mathrm{t}} :: \data{ExtValue}{iso}{} \to \data{Term}{iso}{}\\
    &\compet{v} = \compvt{v}\\
    &\compet{\kw{let}~p_{1}~\kw{=}~\omega~p_{2}~\kw{in}~e_{2}}\\
    &\quad= \kw{let}~p_{1}~\kw{=}~\omega~\comppt{p_{2}}~\kw{in}~\compet{e_{2}}\\
    &\compet{e_{1} + e_{2}} = \compet{e_{1}} + \compet{e_{2}} \\
    &\compet{\alpha{e}} = \alpha\,\compet{e}\\
\end{align*}
\end{minipage}
\caption{The compiler for ISO patterns, pure values, and extended values}
\label{fig:compiler-more}
\end{figure}

The compilers for ISO patterns, pure values, and extended values are shown in \cref{fig:compiler-more}.
We first compile patterns, pure values, and extended values to terms with $\comppt{\_}$, $\compvt{\_}$, and $\compet{\_}$ respectively.
Then we use $\comp{\_}$ to compile the resulting terms to FGGs.
Since patterns, pure values, and extended values are all subsets of terms, the translation from them to terms is straightforward.
We can also prove the following lemma by induction on the typing judgments.
\begin{lemma}\label[lemma]{lemma:trans-iso}
  The compilation from ISO patterns, pure values, and extended values to ISO terms preserves types and semantics:
  \begin{itemize}
  \item If $\TyIsoPattern{\Delta}{p}{a}$, $\TyIsoVal{\Delta}{v}{a}$, and $\TyIsoExt{\Delta}{e}{a}$, then:
    \begin{align}
      \label{eq:trans-iso-more}
      \TyIsoTerm{\Delta}{\comppt{p}}{a}, \quad \TyIsoTerm{\Delta}{\compvt{v}}{a}, \quad \TyIsoTerm{\Delta}{\compet{e}}{a}
    \end{align}
  \item If $\TyIsoPattern{\Delta}{p}{a}$, $\TyIsoVal{\Delta}{v}{a}$, and $\TyIsoExt{\Delta}{e}{a}$, then for any environment $\delta$ of $\Delta$ and semantic value $\val{}$ of $a$:
    \begin{align}
      \label{eq:denot-iso-more}
      \denotpat{\TyIsoPattern{\Delta}{p}{a}}(\delta, \val{}) &= \denot{\TyIsoTerm{\Delta}{\comppt{p}}{a}}(\delta, \val{})\\
      \denotpv{\TyIsoVal{\Delta}{v}{a}}(\delta, \val{}) &= \denot{\TyIsoTerm{\Delta}{\compvt{v}}{a}}(\delta, \val{})\\
      \denotext{\TyIsoExt{\Delta}{e}{a}}(\delta, \val{}) &= \denot{\TyIsoTerm{\Delta}{\compet{e}}{a}}(\delta, \val{})
    \end{align}
  \end{itemize}
\end{lemma}

\section{Proof of \texorpdfstring{\cref{thm:wty-denot}}{Theorem 5.1}}\label{section:proof}

Most cases in \cref{thm:wty-denot} can be proved by induction on the typing judgments.
For \kw{let} terms, we first prove the substitution lemma:
\begin{lemma}\label[lemma]{lemma:subst-typ}
If $\TyIsoTerm{\Gamma}{t_{1}}{a}$ and $\TyIsoTerm{\Delta,x:a}{t_{2}}{b}$, then $\TyIsoTerm{\Gamma;\Delta}{t_{2}[t_{1}/x]}{b}$.
\end{lemma}
This lemma can be proved by induction on $\TyIsoTerm{\Delta,x:a}{t_{2}}{b}$.
Then we can prove the denotational semantics of the substitution:
\begin{lemma}\label[lemma]{lemma:subst-denot}
If $\TyIsoTerm{\Gamma}{t_{1}}{a}$ and $\TyIsoTerm{\Delta,x:a}{t_{2}}{b}$, then for any environment $\delta$ of $\Gamma;\Delta$ and semantic value $\val{}$ of $b$:
\begin{equation}
\denot{\TyIsoTerm{\Gamma;\Delta}{t_{2}[t_{1}/x]}{b}}(\delta, \val{}) = \denot{\TyIsoTerm{\Gamma;\Delta}{\kw{let}~x = t_{1}~\kw{in}~t_{2}}{b}}(\delta, \val{})
\end{equation}
\end{lemma}
This lemma can be proved by induction on $\TyIsoTerm{\Delta,x:a}{t_{2}}{b}$.
We pick the case $t_{2} = x$ below.
Other cases can be proved in a similar way.
\begin{proof}
  Let $t_{2} = x$, then $\Delta = \varnothing$, $t_{2}[t_{1}/x] = t_{1}$.
  Let $\delta$ be an environment of $\Gamma$, $\val{}$ be a value of $a$, we have
  \begin{align}
    &\text{LHS} = \denot{t_{2}[t_{1}/x]}(\delta, \val{}) = \denot{t_{1}}(\delta, \val{})\\
    &\text{RHS} = \denot{\kw{let}~x = t_{1}~\kw{in}~x}(\delta, \val{}) =
      {%
      \sum_{\delta_{2}' \in \denot{x:a}}
      \sum_{\val{}' \in \denot{a}}
      \denot{t_{1}}(\delta, \val{}') \cdot
      \denotpat{x}(\delta_{2}', \val{}') \cdot
      \denot{x}(\delta_{2}', \val{})
      }
  \end{align}
  Only when $\delta_{2}'(x) = \val{}$ and $\val{}' = \val{}$, $\denotpat{x}(\delta_{2}', \val{}') = 1$ and $\denot{x}(\delta_{2}', \val{}) = 1$.
  Therefore, $\text{LHS} = \text{RHS}$.
\end{proof}
\begin{lemma}\label[lemma]{lemma:comp-term}
  The compiler $\data{comp}{}{}$ on ISO terms terminates.
\end{lemma}
\begin{proof}
  Let $\rho$ be a mapping from free variables in a term to $\mathbb{N}$.
  For a variable $x$ not in $\rho$, let $\rho(x) = 1$.
  Define the size of ISO terms as below:
  \begin{align*}
    &\data{|\_|}{\rho}{} ::~ \data{Term}{iso}{} \to \mathbb{N} \\
    &\size{\kw{()}}{\rho} = 1
    &&\size{x}{\rho} = \rho(x)\\
    &\size{\kw{inj}_{l}~t}{\rho} = \size{t}{\rho} + 1
    &&\size{\kw{inj}_{r}~t}{\rho} = \size{t}{\rho} + 1\\
    &\size{\langle{t_{1}, t_{2}}\rangle}{\rho} = \size{t_{1}}{\rho} + \size{t_{2}}{\rho} + 1
    &&\size{\omega~t}{\rho} = \size{t}{\rho} + 1\\
    &\size{\kw{let}~x~\kw{=}~t_{1}~\kw{in}~t_{2}}{\rho} =
      \size{t_{1}}{\rho} + \size{t_{2}}{\rho \cup \{x \mapsto \size{t_{1}}{\rho}\}} + 1
    &&\size{t_{1} + t_{2}}{\rho} = \size{t_{1}}{\rho} + \size{t_{2}}{\rho} + 1\\
    &\size{\kw{let}~p~\kw{=}~t_{1}~\kw{in}~t_{2}}{\rho} = \size{t_{1}}{\rho} + \size{t_{2}}{\rho} + 1
    &&\size{\alpha{t}}{\rho} = \size{t}{\rho} + 1
  \end{align*}
  It can be proved by induction that
  \[\size{t_{2}[t_{1}/x]}{\rho} = \size{t_{2}}{x \mapsto \size{t_{1}}{\rho}}\]
  and
  \[\size{t_{2}[t_1/x]}{\rho} < \size{\kw{let}~x~\kw{=}~t_{1}~\kw{in}~t_{2}}{\rho}\]
  Since the recursive invocations of $\data{comp}{}{}$ always operate on terms of smaller sizes, $\data{comp}{}{}$ terminates.
\end{proof}

\begin{lemma}\label[lemma]{lemma:comp-subst}
  The compilation of substitution preserves semantics: If $\TyIsoTerm{\Gamma}{t_{1}}{a}$ and $\TyIsoTerm{\Delta,x:a}{t_{2}}{b}$, then for any environment $\delta$ of $\Gamma;\Delta$ and semantic value $\val{}$ of $b$:
  \begin{equation}
    \denot{t_{2}[t_{1}/x]}(\delta, \val{}) = \denot{\comp{t_{2}[t_{1}/x]}}_{\gamma}(\compenv{\Gamma;\Delta,\delta}, \compsv{b}{\val{}})
  \end{equation}
\end{lemma}
This lemma can be proved by induction on $\TyIsoTerm{\Delta,x:a}{t_{2}}{b}$.

With all the lemmas above, we can prove \cref{thm:wty-denot}.
We pick the case for the $\kw{let}$ terms.
Other cases can be proved in a similar way.
\begin{proof}
  For the $\kw{let}$ term, by \cref{denot:iso-t-let}, the left hand side has:
  \begin{equation}
    \begin{aligned}
      &\denot{\TyIsoTerm{\Delta_{1};\Delta_{2}}{\kw{let}~p~\kw{=}~t_{1}~\kw{in}~t_{2}}{b}}%
        (\delta_{1} \cup \delta_{2}, \val{}')\\
      &={%
        \sum_{\delta_{2}' \in \denot{\Delta_{2}'}}
        \sum_{\val{} \in \denot{a}}
        \denot{t_{1}}%
        (\delta_{1}, \val{}) \cdot
        \denot{p}%
        (\delta_{2}', \val{}) \cdot
        \denot{t_{2}}%
        (\delta_{2} \cup \delta_{2}', \val{}')
        }
    \end{aligned}
  \end{equation}
  By the definition of $\data{comp}{}{}$ and \cref{e:fgg-einsum}, the right hand side has:
  \begin{align}
    &\denot{\comp{\TyIsoTerm{\Delta_{1};\Delta_{2}}{\kw{let}~p~\kw{=}~t_{1}~\kw{in}~t_{2}}{b}}}_{\gamma}%
      [(\compenv{\Delta_{1};\Delta_{2}, \delta_{1} \cup \delta_{2}}, \compsv{b}{\val{}'})]\notag\\
    &={%
      \begin{aligned}[t]
        &\llbracket\einsum(\comp{t_{1}}[\comp{\Delta_{1}}x_{a}], \comp{p}[\comp{\Delta_{2}'}x_{a}],\\
        &\hphantom{\llbracket\einsum\lparen}\comp{t_{2}}[\comp{\Delta_{2}}\comp{\Delta_{2}'}y_{b}] \to\\
        &\hphantom{\llbracket\einsum\lparen}\comp{\Delta_{1}}\comp{\Delta_{2}}y_{b})\rrbracket_{\gamma}%
          [(\compenv{\Delta_{1};\Delta_{2}, \delta_{1} \cup \delta_{2}}, \compsv{b}{\val{}'})]
      \end{aligned}}\displaybreak[0]\\
    &=
      \sum\nolimits_{%
      \begin{subarray}{l}%
        \delta\in\dom{\comp{\Delta_{1}}x_{a} \cup \comp{\Delta_{2}'}x_{a} \cup \comp{\Delta_{2}}y_{b}} \\
        \delta(\comp{\Delta_{1}}\comp{\Delta_{2}}y_{b})=%
        [(\compenv{\Delta_{1};\Delta_{2}, \delta_{1} \cup \delta_{2}}, \compsv{b}{\val{}'})]%
      \end{subarray}%
      }
      {%
      \begin{array}{l}
        \llbracket{\comp{t_{1}}}\rrbracket_\gamma[\delta(\comp{\Delta_{1}}x_{a})]\cdot{}\\
        \llbracket{\comp{p}}\rrbracket_\gamma[\delta(\comp{\Delta_{2}'}x_{a})]\cdot{}\\
        \llbracket{\comp{t_{2}}}\rrbracket_\gamma[\delta(\comp{\Delta_{2}}\comp{\Delta_{2}'}y_{b})]\\
      \end{array}}\displaybreak[0]\\
    &=
      \sum\nolimits_{%
      \begin{subarray}{l}%
        \delta\in\dom{\comp{\Delta_{2}'}x_{a}} \\
      \end{subarray}%
      }
      {%
      \begin{array}{l}
        \llbracket{\comp{t_{1}}}\rrbracket_\gamma[(\compenv{\Delta_{1}, \delta_{1}},\delta(x_{a}))]\cdot{}\\
        \llbracket{\comp{p}}\rrbracket_\gamma[(\delta(\comp{\Delta_{2}'}),\delta(x_{a}))]\cdot{}\\
        \llbracket{\comp{t_{2}}}\rrbracket_\gamma[(\compenv{\Delta_{2},\delta_{2}},\delta(\comp{\Delta_{2}'}),\compsv{b}{\val{}'})]\\
      \end{array}
      }\displaybreak[0]\\
    \label{pf:let-last-term}
    &=
      \sum\nolimits_{\delta'\in\dom{\comp{\Delta_{2}'}}}
      \sum\nolimits_{\delta''\in\dom{x_{a}}}
      {%
      \begin{array}{l}
        \llbracket{\comp{t_{1}}}\rrbracket_\gamma[(\compenv{\Delta_{1}, \delta_{1}},\delta''(x_{a}))]\cdot{}\\
        \llbracket{\comp{p}}\rrbracket_\gamma[(\delta'(\comp{\Delta_{2}'}),\delta''(x_{a}))]\cdot{}\\
        \llbracket{\comp{t_{2}}}\rrbracket_\gamma[(\compenv{\Delta_{2},\delta_{2}},\delta'(\comp{\Delta_{2}'}),\compsv{b}{\val{}'})]\\
      \end{array}
      }
  \end{align}
  To continue, we use the fact that there is a bijection between $\denot{\Delta_{2}'}$ and $\dom{\comp{\Delta_{2}'}}$, namely, for every $\delta' \in \denot{\Delta_{2}'}$ we have $\{\comp{\Delta_{2}'} \mapsto \compenv{\Delta_{2}',\delta'}\} \in \dom{\comp{\Delta_{2}'}}$.
  Moreover, there is a bijection between $\denot{a}$ and $\dom{x_a}$, namely, for every $\val{} \in \denot{a}$ we have $\{x_a \mapsto \compsv{a}{\val{}}\} \in \dom{x_a}$.
  Thus, \cref{pf:let-last-term} can be rewritten as:
  {
  \allowdisplaybreaks
  \begin{align}
    &\sum\nolimits_{\delta'\in\denot{\Delta_{2}'}}
      \sum\nolimits_{\begin{subarray}{l}\val{}\in\denot{a}\end{subarray}}
      {%
      \begin{array}{l}
        \llbracket{\comp{t_{1}}}\rrbracket_\gamma[(\compenv{\Delta_{1}, \delta_{1}},\compsv{a}{\val{}})]\cdot{}\\
        \llbracket{\comp{p}}\rrbracket_\gamma[(\compenv{\Delta_{2}', \delta'},\compsv{a}{\val{}})]\cdot{}\\
        \llbracket{\comp{t_{2}}}\rrbracket_\gamma[(\compenv{\Delta_{2},\delta_{2}},\compenv{\Delta_{2}', \delta'},\compsv{b}{\val{}'})]\\
      \end{array}
      }\\
    &=
      \sum\nolimits_{\delta'\in\denot{\Delta_{2}'}}
      \sum\nolimits_{\begin{subarray}{l}\val{}\in\denot{a}\end{subarray}}
      {%
      \begin{array}{l}
        \llbracket{\comp{t_{1}}}\rrbracket_\gamma[(\compenv{\Delta_{1}, \delta_{1}},\compsv{a}{\val{}})]\cdot{}\\
        \llbracket{\comp{p}}\rrbracket_\gamma[(\compenv{\Delta_{2}', \delta'},\compsv{a}{\val{}})]\cdot{}\\
        \llbracket{\comp{t_{2}}}\rrbracket_\gamma[(\compenv{\Delta_{2};\Delta_{2}',\delta_{2} \cup \delta'},\compsv{b}{\val{}'})]\\
      \end{array}
    }\\
    &=
      \sum\nolimits_{\delta'\in\denot{\Delta_{2}'}}
      \sum\nolimits_{\begin{subarray}{l}\val{}\in\denot{a}\end{subarray}}
      \llbracket{t_{1}}\rrbracket(\delta_{1}, \val{})\cdot
      \llbracket{p}\rrbracket(\delta',\val{})\cdot
      \llbracket{t_{2}}\rrbracket(\delta_{2} \cup \delta',\val{}')
      \text{\qquad(by IH)}
  \end{align}
  Therefore,
  \begin{align}
    &\denot{\TyIsoTerm{\Delta_{1};\Delta_{2}}{\kw{let}~p~\kw{=}~t_{1}~\kw{in}~t_{2}}{b}}%
    (\delta_{1} \cup \delta_{2}, \val{}')\notag\\
    &=\denot{\comp{\TyIsoTerm{\Delta_{1};\Delta_{2}}{\kw{let}~p~\kw{=}~t_{1}~\kw{in}~t_{2}}{b}}}_{\gamma}%
      [(\compenv{\Delta_{1};\Delta_{2}, \delta_{1} \cup \delta_{2}}, \compsv{b}{\val{}'})]
  \end{align}
  }
\end{proof}

\section{Proof of \texorpdfstring{\cref{thm:einsum-expand}}{Theorem 6.1}}\label{section:proof-einsum}

\begin{proof}[Proof sketch]
The denotation on the left hand side of \labelcref{e:einsum-expand} is a summation over indices of shape $(\vec{x}\vec{y}\cup\vec{z}) \setminus \vec{w}$.
This shape turns out to be the disjoint union of two shapes: $\vec{v}$ consists of the dimensions that are not used to index any input tensor, and $(\vec{x}\cup\vec{y}) \setminus \vec{u}$ consists of the dimensions that are used to index some input tensors.
The main argument, then, is to split the summation into two nested summations.
One of them ranges over indices of shape~$\vec{v}$, and reduces to the constant factor~$f$ because $\vec{v}$ is unused.

Specifically, for any index tuple~$I$ such that $\{\vec{w} \mapsto I\}$, using the conditions $\vec{y} \disjoint \vec{z}$ and $\vec{w} \sqsubseteq \vec{x}\cup\vec{y}\vec{z}$, the two sides of \labelcref{e:einsum-expand} separately reduce to
\begin{multline}
\label{e:einsum-expand-split}
      \sum_{
        \delta'\in\dom{\vec{v}}
      }
      \sum_{
        \delta\in\dom{(\vec{x}\cup\vec{y})\setminus\vec{u}}
      }
\begin{array}[t]{@{}l@{}}
        \llbracket F \rrbracket_\gamma [(\delta'\cup\delta\cup\{\vec{w}\mapsto I\})(\vec{x})] \\[.5\jot]\enspace{}\cdot
        \llbracket E \rrbracket_\gamma [(\delta'\cup\delta\cup\{\vec{w}\mapsto I\})(\vec{y})]
\end{array}
\\=
    \Bigl(\prod_{i \in \vec{v}}|i|\Bigr) \cdot
    \sum_{
        \delta\in\dom{(\vec{x}\cup\vec{y})\setminus\vec{u}}
      }
\begin{array}[t]{@{}l@{}}
    \llbracket F \rrbracket_\gamma [(\delta\cup\{\vec{w}\mapsto I\})(\vec{x})] \\[.5\jot]\enspace{}\cdot
    \llbracket E \rrbracket_\gamma [(\delta\cup\{\vec{w}\mapsto I\})(\vec{y})]
\end{array}
\end{multline}
  Since $\vec{v}$ doesn't contain dimensions from $\vec{x}$ or $\vec{y}$, we know $\vec{v}$ is not used to index the denotations of $F$ and $E$.
  Thus, the index function $\delta'$, whose domain is $\vec{v}$, is not used in the first summation in \labelcref{e:einsum-expand-split}.
  Therefore, that summation reduces to the constant factor $\prod_{i \in \vec{v}}|i|$ on the right hand side.
\end{proof}

\end{document}